\documentclass[11pt]{article}

\usepackage{graphicx}
\usepackage{multicol}
\usepackage{color}
\usepackage{epsfig,latexsym}

\definecolor{verdon}{cmyk}{1,0.5,1,0}
\definecolor{blue}{cmyk}{0.8,0.8,0,0.}
\definecolor{red}{cmyk}{0.2,1,1,0.0}
\newcommand{\cita}[1]{{\color{blue} \cite{#1}}}

\def\lapprox{\mathrel{\mathop  {\hbox{\lower0.5ex\hbox{$\sim$}
\kern-1.1em\lower-0.7ex\hbox{$<$}}}}}
\def\gapprox{\mathrel{\mathop  {\hbox{\lower0.5ex\hbox{$\sim$}
\kern-1.1em\lower-0.7ex\hbox{$>$}}}}}

\begin{document}

\title{\color{verdon} Constraints on the opacity profile of the sun from helioseismic observables and solar neutrino flux measurements}
\author{F.L. Villante$^{1}$
\vspace{0.5 cm}\\
{\small\em $^{1}$Universit\`a dell'Aquila and INFN - LNGS, L'Aquila - Italy}}

\date{}

\maketitle

\def\abstractname{\color{red}\bf Abstract}
\begin{abstract}
{\footnotesize Motivated by the solar composition problem and by using the recently developed Linear Solar Model approach \cite{noi},
we analyze the role of opacity and metals in the sun. After a brief discussion of the relation  between the effects produced by a variation of composition
and those produced by a modification of the radiative opacity, we 
calculate numerically the opacity kernels that, in a linear approximation, 
relate an arbitrary opacity variation to the corresponding modification 
of the solar observable properties. We use these opacity kernels
to  discuss the present  constraints on opacity (and composition) provided by helioseismic and solar neutrino data.}
\end{abstract}

\newpage

\section{Introduction}

\begin{figure}[t]
\par
\begin{center}
\includegraphics[width=8.5cm,angle=0]{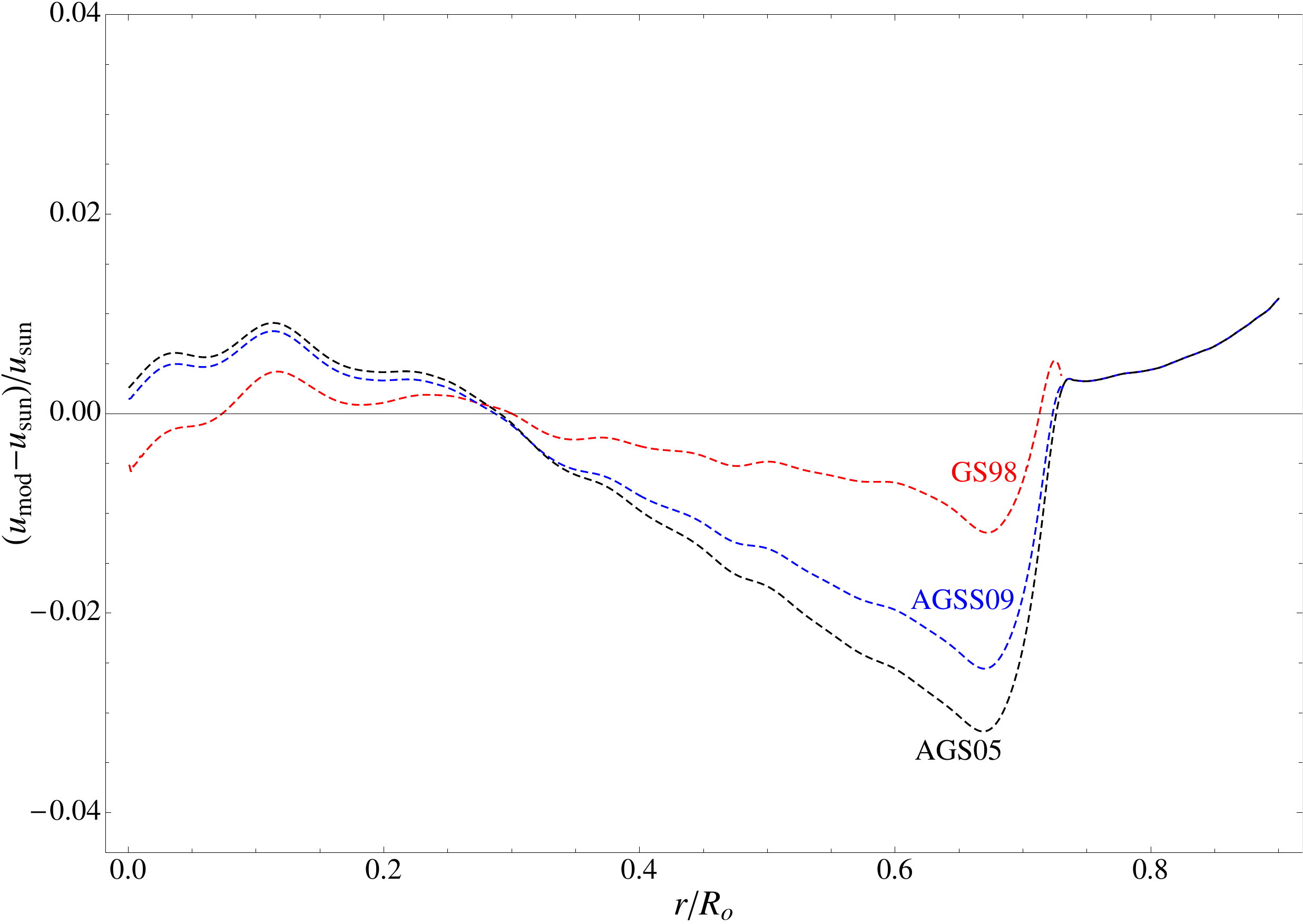}
\includegraphics[width=8.5cm,angle=0]{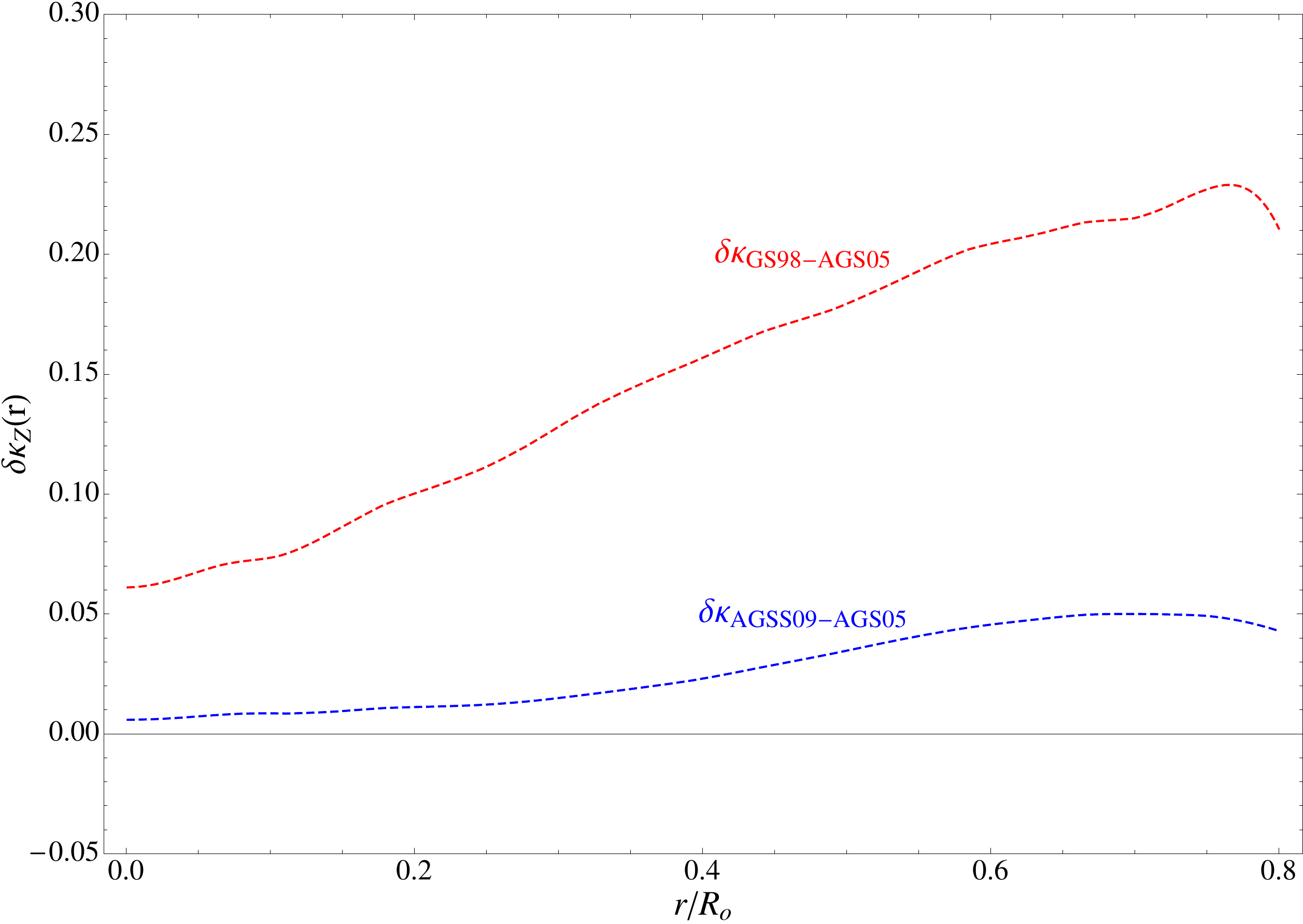}
\end{center}
\par
\vspace{-5mm} \caption{\em {\protect\small  Left Panel: The difference between the squared isothermal sound speed inferred from helioseismic data
and the predictions of solar models implementing AGS05 (black), GS98 (red) and AGSS09 (blue) heavy element admixtures. 
Right panel: The composition opacity changes $\delta \kappa_{Z}(r)$ that corresponds to using the GS98 (red) and the AGSS09 (blue)
composition in place of the AGS05 composition.}}
\label{fig1}
\end{figure}

In the last few years a new solar problem has emerged. Recent determinations 
of the photospheric heavy element abundances \cite{as05,as09,caffau09} 
indicate that the sun metallicity is lower than previously
assumed \cite{gs98}. Solar models that incorporate these lower abundances are no more able to
reproduce the helioseismic results. As an example, the sound speed predicted by standard solar models (SSMs) 
implementing the heavy element admixture of \cite{as05} disagrees at the bottom of the convective envelope by
$\sim 10 \, \sigma$ with the value inferred by helioseismic data (see e.g. \cite{BBPS05} and black dashed line in fig.\ref{fig1}). In addition, 
the predicted surface helium abundance is lower by $\sim 6 \, \sigma$ and the radius 
of the convective envelope is larger by $\sim 15 \, \sigma$ with respect to the helioseismic results.
 Detailed studies have been done to resolve this controversy, see e.g.\cita{basu}.
The latest determinations of the solar photospheric composition \cite{as09,caffau09} alleviate the discrepancies
but a definitive solution of the ``solar composition problem'' still has to be obtained.

 The main effect of changing the heavy element admixture is to modify 
the opacity profile of the sun. It is, thus, evident that the comprehension of the solar 
composition problem is intimately related to understanding the role of opacity 
in solar  modelling. Several authors have investigated the effects of opacity
changes on the solar structure by using different methods and assumptions
(see e.g. \cite{chris} and references therein). 
Here, we continue their work, completing and extending the analysis of \cite{chris} by using 
a different and original approach.
The final goal is to provide the instruments to analyze in transparent and efficient way 
the role of the opacity in the sun
and to perform a critical ``step-by-step'' discussion of the present 
constraints on opacity (and composition) provided by observable properties of the sun.

In order to calculate the effects of arbitrary opacity changes on the sun, we use the linear solar model (LSM) approach,
presented in \cita{noi, noi2} and briefly summarized in the appendix. In this approach, 
the structure equations of the present sun are linearized and, by estimating the (variation of) 
the present solar composition from the (variation of) the nuclear reaction rates and 
elemental diffusion efficiency in the present sun, we obtain a linear system of ordinary 
differential equations that can be easily solved and that completely determines the physical 
and chemical properties of the sun. It was shown in \cita{noi} that this kind of approach reproduces
with good accuracy the results of non-linear evolutionary solar models
and, thus, can be used to study the role of parameters 
and assumptions in solar model construction. 

By considering localised opacity changes in LSM approach, 
we determine numerically the kernels that, in a linear approximation, 
relate an arbitrary opacity variation to the corresponding modification 
of the solar observable properties.
These opacity kernels are useful in several respects.
First, they allow us to individuate the region of the sun whose opacity is probed with maximal 
sensitivity by each observable quantity.
Then, they permit us to show that effects produced by variations of opacity in different 
region of the sun can compensate among each others.
Finally, they will be used to discuss how the
different pieces of observational information cooperate 
to determine the present constraints on the opacity profile of the sun. 

The plan of the paper is the following.
In sect.~\ref{sectII}, we discuss the relation between the effects produced by a variation of the heavy element admixture 
and those produced by a modification of the radiative opacity. We show that the relevant quantity
is the variation of the opacity profile $\delta \kappa(r)$ defined in eq.~(\ref{kappasource}), that is approximately  given by 
the superposition of the {\em intrinsic} opacity change $\delta \kappa_{\rm I}(r)$ and 
the {\em composition} opacity change $\delta \kappa_{\rm Z}(r)$, 
defined in eqs.~(\ref{kappaintrinsic}) and (\ref{kappacomposition}) respectively.
In sect.~\ref{sectIII}, we define the opacity kernels
and we describe the 
method adopted to calculate them.
In sect.~\ref{sectIV}, we calculate numerically 
the kernels for the squared isothermal sound speed $u(r)\equiv P(r)/\rho(r)$, the surface helium abundance $Y_{\rm b}$, the radius of the convective
envelope $R_{\rm b}$ and the various neutrino fluxes $\Phi_{\nu}$. Moreover,
we discuss the constraints on $\delta \kappa(r)$ provided by the present observational data. 
Finally, in sect.~\ref{sectVIII} we summarize our results.
A conclusive view of the constraints on $\delta \kappa(r)$  is provided by fig.~\ref{FigFinal}.



\section{The relation between opacity and metals}
\label{sectII}

We consider a modification of the opacity $\kappa(\rho,T,Y,Z_{\rm i})$ and/or of the heavy element  photospheric 
admixture $\left\{z_{\rm i}\right\}$, expressed here in terms of the quantities $z_{\rm i} \equiv Z_{\rm i,b} /X_{\rm b}$
where $Z_{\rm i, b}$ is the surface abundance of the $i-$element  and  $X_{\rm b}$ is that of hydrogen. 
If we neglect the role of metals in the equation of state and in the energy generation coefficient, 
the only effect of these changes is to modify radiative energy transport in the sun. 
The relevant parameter, in this respect, is the total variation 
of the opacity in the shell $r$ of the present sun, given by:
\begin{equation}
\delta \kappa^{\rm tot}(r) = \frac{\kappa(\rho(r),T(r),Y(r),Z_{\rm i}(r))}
{\overline{\kappa}(\overline{\rho}(r),\overline{T}(r),\overline{Y}(r),\overline{Z}_{\rm i}(r))} -1
\end{equation}
where $\overline{\kappa}(\overline{\rho}(r),\overline{T}(r),\overline{Y}(r),\overline{Z}_{\rm i}(r))$ is the opacity 
profile of the SSM, while $\kappa(\rho(r),T(r),Y(r),Z_{\rm i}(r))$ is the opacity profile 
of the solar model that implements the modified opacity and photospheric composition\footnote{
The notation $\overline{Q}$ indicates, here and in the following, the SSM's value 
for the generic quantity Q.}.
We note that the quantity $\delta \kappa^{\rm tot}(r)$ is not related in a direct way to the performed 
variations of opacity and composition.
In order to calculate it, we have to take into account 
that the ``perturbed'' sun 
has different density ($\rho$), 
temperature ($T$) and chemical composition profiles ($Y$ and $Z_{\rm i}$) with respect to the SSM.
These are not known {\em a priori} but have to be obtained as a result of 
numerical solar modelling.

A relevant simplification is obtained in the  LSM approach presented in ref.\cite{noi},
where one assumes that: {\em i)} the performed changes of opacity and heavy element admixture are small;
{\em ii)} the (variation of) the chemical composition of the sun can be estimated from the 
(variation of) nuclear reaction rate and diffusion efficiency of the present sun; 
{\em iii)} the variation of the metal admixture has a negligible {\em direct} effect
on nuclear production of helium and on diffusion efficiency. 
In this case, the relation between $\delta \kappa^{\rm tot}(r)$ and 
the performed variation of 
opacity and admixture 
can be worked out explicitly,
as it is described in the appendix.
It can be shown, moreover, that the source term  $\delta \kappa(r)$ that 
drives the modification of the solar properties\footnote{In mathematical terms, 
the quantity $\delta \kappa(r)$ represents the inhomogeneous term in the linearized structure equation of the present 
sun.}
and that can be constrained by observational data 
can be written as the sum of two contributions:
\begin{equation}
\delta \kappa(r) = \delta \kappa_{\rm I}(r) + \delta \kappa_{\rm Z}(r)
\label{kappasource}
\end{equation}
The first term $\delta \kappa_{\rm I}(r)$, which we refer to as {\em intrinsic} opacity change, 
represents the fractional variation of the opacity  along the SSM profile and it is given by:
\begin{equation}
\label{kappaintrinsic}
\delta \kappa_{\rm I}(r) = \frac{\kappa(\overline{\rho}(r),\overline{T}(r),\overline{Y}(r),\overline{Z}_{i}(r))}
{\overline{\kappa}(\overline{\rho}(r),\overline{T}(r),\overline{Y}(r),\overline{Z}_{i}(r))} -1
\end{equation}
This contribution is obtained when we revise the opacity function $\kappa(\rho,T,Y,Z_{\rm i})$
and/or we introduce new effects, like e.g. the accumulation of few GeVs WIMPs in the solar 
core (see e.g.\cite{WimpsNoi} and references therein) that mimics a decrease of the opacity at the solar center.

The second term $\delta \kappa_{\rm Z}(r)$, 
which we refer to as {\em composition} opacity change, 
describes the effects of a variation of $\left\{ z_{\rm i} \right\}$.
It takes into account that a modification of the photospheric admixture implies a different 
distribution of metals inside the sun and, thus, a different opacity profile, even if the function
$\kappa(\rho,T,Y,Z_{\rm i})$ is unchanged. The contribution $\delta \kappa_{\rm Z}(r)$ is given by
(see appendix A and B):
\begin{equation}
\delta \kappa_{\rm Z}(r) = \frac{
\overline{\kappa}(\overline{\rho}(r),\overline{T}(r),\overline{Y}(r),Z_{i}(r))}
{\overline{\kappa}(\overline{\rho}(r),\overline{T}(r),\overline{Y}(r),\overline{Z}_{i}(r))} - 1 
\end{equation}
where $Z_{\rm i}(r) = \overline{Z}_{\rm i}(r) \, (z_{\rm i} / \overline{z}_{\rm i})$ and can be calculated as:
\begin{equation}
\label{kappacomposition}
\delta \kappa_{\rm Z}(r)  \simeq \sum_{i} \left.\frac{\partial \ln \overline {\kappa}}{\partial \ln Z_{i}} \right|_{\rm SSM} \; \delta z_{\rm i,b}
\end{equation} 
where $\delta z_{\rm i}$ represents the fractional variation of $z_{\rm i}$  and 
the symbol $|_{\rm SSM}$ indicates that we calculate the derivatives along the density, temperature and 
chemical composition profiles predicted by the SSM.

\begin{table}[t]
\begin{center}{
\begin{tabular}{l|cc|ccccc}
& $\delta R_{\rm b}$  &  $\Delta Y_{\rm b}$ &
$\delta\Phi_{\rm pp}$ & $\delta\Phi_{\rm Be}$    & $\delta\Phi_{\rm B}$ & $\delta\Phi_{\rm O}$ & $\delta\Phi_{\rm N}$   \\
\hline
\hline
GS98 - AGS05    & -0.019  & 0.019      & -0.011   & 0.14   & 0.26   & 0.56 & 0.49   \\
AGSS09 - AGS05 & -0.0056 & 0.0028  & -0.0014 & 0.018 & 0.034 &  0.13 & 0.12  \\
\hline
\end{tabular}
}\end{center}\vspace{0.4cm} \caption{\em {\protect\small  
The variations of helioseismic and solar neutrino observables produced by a variation of
 the heavy element admixture. The above results have been estimated within the LSM approach,
by applying the opacity changes $\delta \kappa_{\rm Z}(r)$ shown in the right panel of fig.\ref{fig1}.
For CNO neutrinos, we also considered that the fluxes scales proportionally to the total CN-abundance.
Note that the absolute variation is reported for the surface helium abundance,
whereas the relative variations are shown for all the other quantities. 
\label{tab1} 
}}\vspace{0.4cm}
\end{table}

Eq.(\ref{kappasource}), although being approximate, is quite useful because it  makes explicit the connection (and the degeneracy) between the effects produced by
a modification of the radiative opacity and of those produced by a modification of the heavy element admixture.
In this paper, we take  as a reference the Asplund, Grevesse, Sauval 2005 composition (AGS05) \cite{as05}
and we refer with 'SSM predictions' to the numerical results obtained by using this composition as input for evolutionary
solar model calculations\footnote{The black dashed lines in the left panel of Fig.1 has been obtained by using the FRANEC code.
See \cite{Franec}  for a description of the code  and \cite{noi} for a description of the results.}. Other compilation
can be considered, like e.g. the Grevesse, Sauval 1998 (GS98) \cite{gs98}  
or the more recent Asplund, Grevesse, Sauval, Scott 2009 (AGSS09) \cite{as09}. The red and blue dashed lines in 
the right panel of Fig.\ref{fig1} correspond to the opacity change $\delta \kappa_{\rm Z}(r)$ that are obtained 
when we use the GS98 and the AGSS09 admixture in place of the AS05 composition, as calculated by applying rel.(\ref{kappacomposition}) 
and by using the logarithmic derivatives $\partial \ln \kappa/\partial \ln Z_{\rm i}$ presented in 
Fig.12 of \cite{basu}. We observe that the variation from AGS05 to GS98 (AGSS09) heavy element admixture 
corresponds to increasing the opacity by about 5\% (1\%) at the center of the sun and by about 20\% (5\%) at the bottom of the
convective region. The effects of these opacity changes on helioseismic and solar neutrino observables, calculated in the 
LSM approach,  are described in tab.\ref{tab1} and in the left panel of fig.\ref{fig1}. They can be compared with 
the results of full non-linear evolutionary codes reported e.g. in \cite{serenelli}, obtaining a satisfactory agreement.

\section{The method}
\label{sectIII}

In the following, we consider the effects produced by a generic variation of the opacity profile
$\delta \kappa(r)$, without discussing whether this is due to a change of the function
$\kappa(\rho,T,Y,Z_{\rm i}$) or to a change of the admixture $\left\{ z_{\rm i} \right\}$.  As a results of this
modification, we obtain a solar model 
that deviates from SSM predictions. If the opacity variation is sufficiently small (i.e. $\delta \kappa(r) \ll 1$), the sun 
responds linearly. In this case, the fractional variation of a generic quantity $Q$, defined as:
\begin{equation}
\delta  Q = \frac{Q}{\overline{Q}}-1
\end{equation}
can be related to $\delta \kappa(r)$ by the linear relation:
\begin{equation}
\delta Q = \int dr \; K_{Q}(r) \; \delta \kappa(r)
\label{linear}
\end{equation}
The kernel $K_{Q}(r)$ represents the functional derivative with 
respect to opacity and allows to quantify the sensitivity of 
$Q$  to opacity variations in different zones of the sun.

In this paper, we determine numerically the kernels $K_{Q}(r)$ for 
helioseismic observables and solar neutrino fluxes, by using the linear solar model (LSM) approach
presented in \cita{noi, noi2} and briefly summarized in the appendix.
In this approach, 
the structure equations of the present sun are linearized and, by estimating the (variation of) 
the present solar composition from the (variation of) the nuclear reaction rates and 
elemental diffusion efficiency in the present sun, we obtain a linear system of ordinary 
differential equations that can be easily solved and that completely determines the physical 
and chemical properties of the sun. It was shown in \cita{noi} that this kind of approach reproduces
with good accuracy the results of non-linear evolutionary solar models
and, thus, can be used to study in an efficient and transparent way the role of parameters 
and assumptions in solar model construction\footnote{The results presented in this paper can be compared with 
those presented in \cite{chris} where some of the
kernels presented here are calculated by using static solar models and/or evolutionary models with simplified 
equation of state and without elemental diffusion. Where comparison is possible, a very good agreement is achieved
showing that the linearization procedure adopted here and the simplifying assumptions implied in 
\cite{chris} do not introduce relevant errors. In order to make the comparison, one should note that the definition of the kernels given in 
\cite{chris} differs form that adopted in this paper.}.

 The estimate of $K_{Q}(r)$ at a given point $r=r_0$ is obtained by performing a localized increase of opacity
in the vicinity of $r_0$. More precisely, we calculate the variation $\delta Q(r_0)$ produced by a normalized\footnote{We remark that 
the LSM are linear ``by construction''. The validity of the the linear relation (\ref{linear}) is, thus, not limited by 
the condition $\delta \kappa(r)\ll1$.} gaussian increase of opacity centered in $r_0$:
\begin{equation}
\delta \kappa(r) = G(r-r_0)\equiv \frac{1}{\sqrt{2\pi}\delta r}\exp\left[-\frac{(r-r_0)^2}{2\delta r^2}\right]
\label{gaussianincrease}
\end{equation}
with $\delta r = 0.01 R_{\odot}$, and we assume that:
\begin{equation}
K_{Q}(r_{0}) = \delta Q({r_0})
\end{equation}
This corresponds to the approximation:
\begin{equation}
K_{Q}(r_{0})\simeq \int dr \; K_{Q}(r) \; G(r-r_{0})
\end{equation}
which is adequate to describe all the situations in which we consider opacity variations $\delta \kappa (r)$
which vary on scale larger than $\delta r = 0.01 R_{\odot}$.


\begin{figure}[t]
\par
\begin{center}
\includegraphics[width=8.5cm,angle=0]{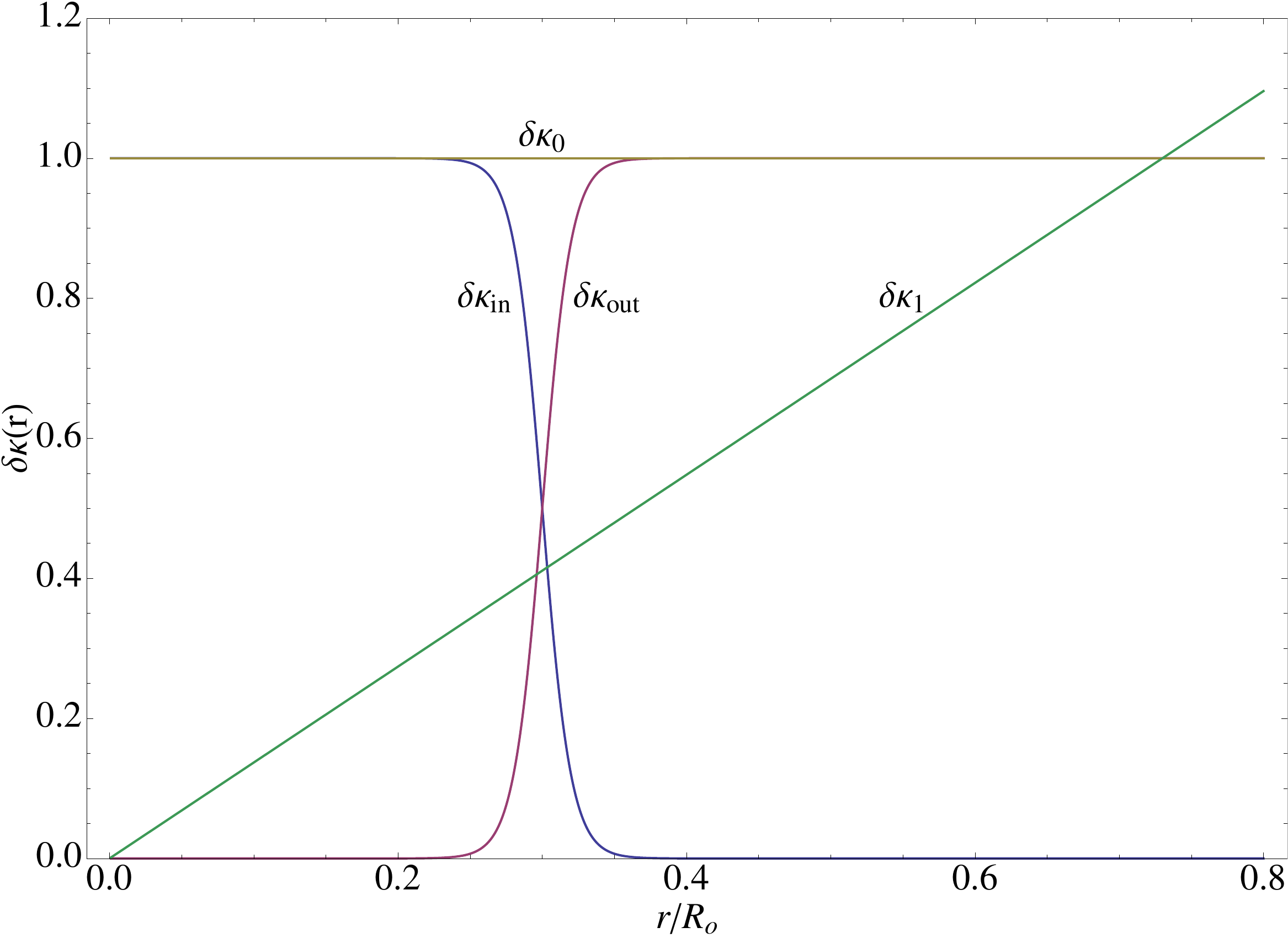}
\end{center}
\par
\vspace{-5mm} \caption{\em {\protect\small  The functions $\delta \kappa_{\rm in}(r)$, $\delta \kappa_{\rm out}(r)$, $\delta \kappa_{0}(r)$ and  $\delta \kappa_{1}(r)$
adopted in the parametrizations (\ref{twozones},\ref{lineartilt}) and defined in eqs.(\ref{param-def}).}}
\label{figKappa}
\end{figure}


The kernels $K_{Q}(r)$ can be used to calculate the effects of an arbitrary opacity change $\delta \kappa(r)$
and allow us to discuss the role of opacity in general terms. In order to consider specific situations and to
understand what kind of experimental constraints are provided by each observable 
quantity $Q$, it is useful, however, to consider the simple parametrizations:
\begin{eqnarray}
\label{twozones}
\delta \kappa(x) &=& A_{\rm in} \; \delta \kappa_{\rm in}(x) + A_{\rm out} \; \delta \kappa_{\rm out}(x)\\
\label{lineartilt}
\delta \kappa(x) &=& A_{0} \; \delta \kappa_{0}(x) + A_{1} \; \delta \kappa_{1}(x)
\end{eqnarray}
with $A_{\rm in}$, $A_{\rm out}$, $A_{0}$ and $A_{1}$ free adjustable parameters and:
\begin{eqnarray}
\label{param-def}
\nonumber
\delta \kappa_{\rm in} (r) &\equiv& \frac{1}{\exp{\left[(r-r_{\rm c})/A \right]}+1}\\
\nonumber
\delta \kappa_{\rm out} (r) &\equiv& 1 - \frac{1}{\exp{\left[(r-r_{\rm c})/A \right]}+1} \\
\nonumber
\delta \kappa_{\rm 0} (r) &\equiv& 1\\
\delta \kappa_{\rm 1} (r) &\equiv& \frac{r}{\overline{R}_{\rm b}}
\end{eqnarray}
where $r_{\rm c}=0.3 R_{\odot}$, $A=0.01$, while $\overline{R}_{\rm b} = 0.730 R_{\odot}$
is the radius of the convective envelope predicted by SSM (see \cite{noi}).
The functions $\delta \kappa_{\rm in}(r)$ and $\delta \kappa_{\rm out}(r)$ correspond to a constant 
increase of the opacity in the energy producing zone ($r\le 0.3 R_{\odot}$) and in 
the outer radiative region ($r\ge 0.3 R_{\odot}$), see Fig.\ref{figKappa}. 
They have been defined in such a way that $\delta \kappa_{\rm in} (r) + \delta \kappa_{\rm out} (r)\equiv 1$. 
The function $\delta \kappa_{0}(r)$ and $\delta \kappa_{1}(r)$ correspond to a global rescaling and to a linear tilt
of the opacity profile.

In linear approximation, the fractional variation $\delta Q$ 
produced by the opacity profiles (\ref{twozones},\ref{lineartilt}) can be expressed as:
\begin{eqnarray}
\nonumber
\delta Q &=& A_{\rm in} \; \delta Q_{\rm in} + A_{\rm out} \; \delta Q_{\rm out}\\
\label{Q-response}
\delta Q &=& A_{0} \; \delta Q_{0} + A_{1} \; \delta Q_{1}
\end{eqnarray}
where the coefficients $\delta Q_{j}$ are given by:
\begin{equation}
\delta Q_{j} = \int dr \; K_{Q}(r) \; \delta \kappa_{j}(r)
\end{equation}
with $j = {\rm in},\, {\rm out},\, 0, \,1$. We report these coefficients in Tab.\ref{tab2} for helioseismic observables
and solar neutrino fluxes.

\begin{table}[t]
\begin{center}{
\begin{tabular}{c|cccc}
$\delta Q$ &  $\delta Q_{\rm in}$ & $\delta Q_{\rm out}$
&  $\delta Q_{\rm 0}$ & $\delta Q_{\rm 1}$\\
\hline
\hline
$\delta u(0.1 R_{\odot})$ &  0.019 & -0.036 & -0.017 & -0.022 \\
$\delta u(0.2 R_{\odot})$ &  0.052 & -0.054 & -0.0025  & -0.014 \\
$\delta u(0.4 R_{\odot})$ &  -0.084 & 0.087  & 0.0037  &  0.031\\
$\delta u(0.65 R_{\odot})$ &  -0.16 & 0.17 & 0.011  & 0.11 \\
\hline
$\Delta Y_{\rm b}$ & 0.073  & 0.069 & +0.142  &  0.062\\
$\delta R_{\rm b}$ &  0.12 & -0.14 & -0.02  & -0.10 \\
\hline
$\delta\Phi_{\rm pp}$ & -0.069 &  -0.031 & -0.100 & -0.030 \\
$\delta\Phi_{\rm Be}$ & 0.85 & 0.41 & 1.26 & 0.38 \\
$\delta\Phi_{\rm B}$ & 1.93 & 0.75  & 2.68 &  0.68 \\
$\delta\Phi_{\rm O}$ & 1.65 & 0.50 & 2.15 & 0.48 \\
$\delta\Phi_{\rm N}$ & 1.14 & 0.28 & 1.43 & 0.30 \\
\hline
\end{tabular}
}\end{center}\vspace{0.4cm} \caption{\em {\protect\small  The coefficients $\delta Q_{\rm in}$, $\delta Q_{\rm out}$, $\delta Q_{0}$ and  $\delta Q_{1}$
defined in eq.(\ref{Q-response}), which allow to calculate the response of helioseismic and solar neutrino observables to opacity changes parameterized by eqs.(\ref{twozones},\ref{lineartilt}).
\label{tab2} 
}}\vspace{0.4cm}
\end{table}

\section{The opacity kernels}
\label{sectIV}

We calculate the opacity kernels for following observable quantities:
the squared isothermal sound speed $u(r)\equiv P(r)/\rho(r)$; 
the surface helium abundance $Y_{\rm b}$;
the depth $R_{\rm b}$ of the convective envelope; 
the solar neutrino fluxes $\Phi_{\nu}$, where the index $\nu = {\rm pp, Be, B, N, O}$
refers to the neutrino producing reactions according to the usual convention.
Our results are presented in the following sub-sections, together   
with a discussion of the present observational 
constraints on $\delta \kappa(r)$.

\subsection{The sound speed}

\begin{figure}[t]
\par
\begin{center}
\includegraphics[width=8.5cm,angle=0]{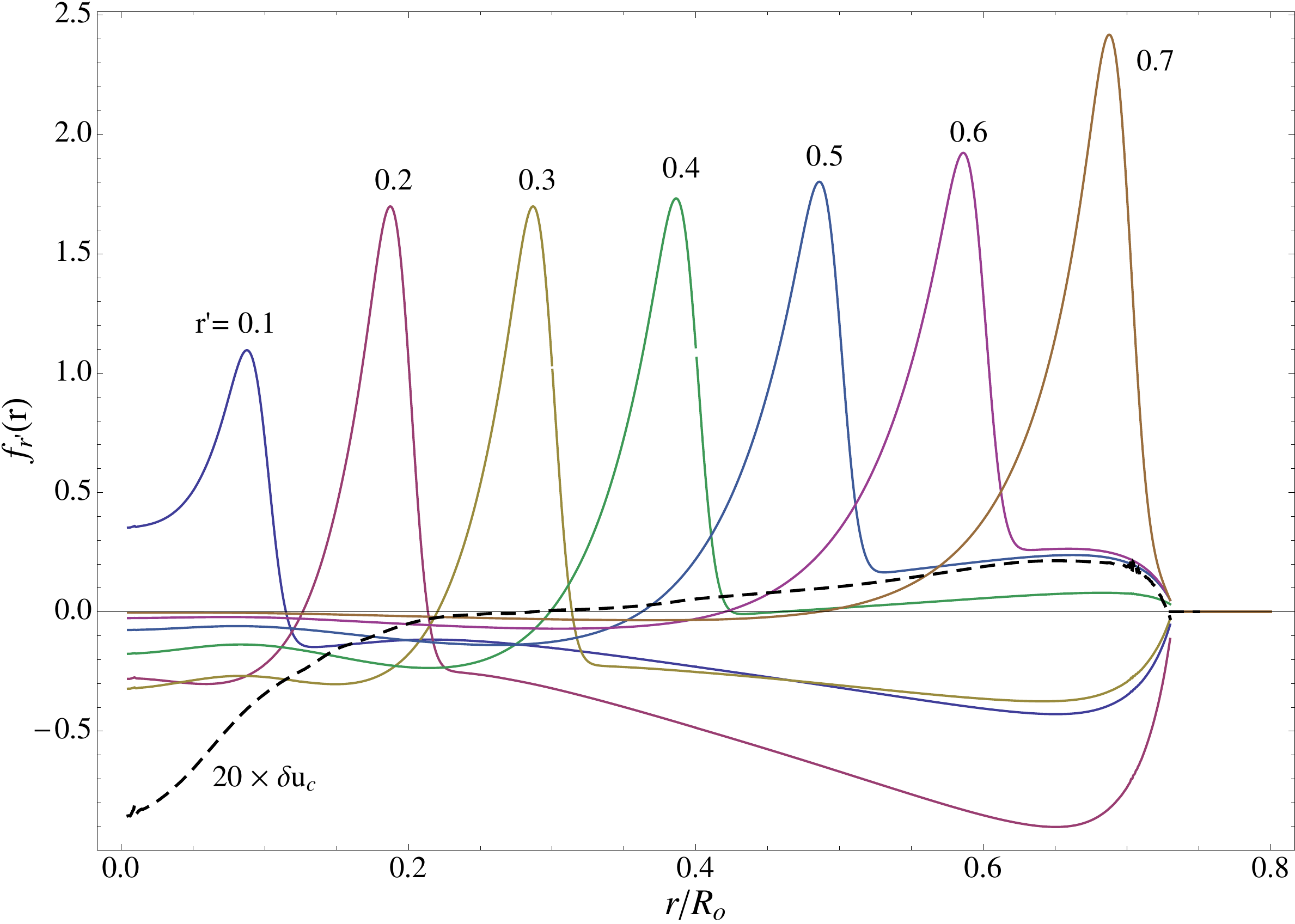}
\includegraphics[width=8.5cm,angle=0]{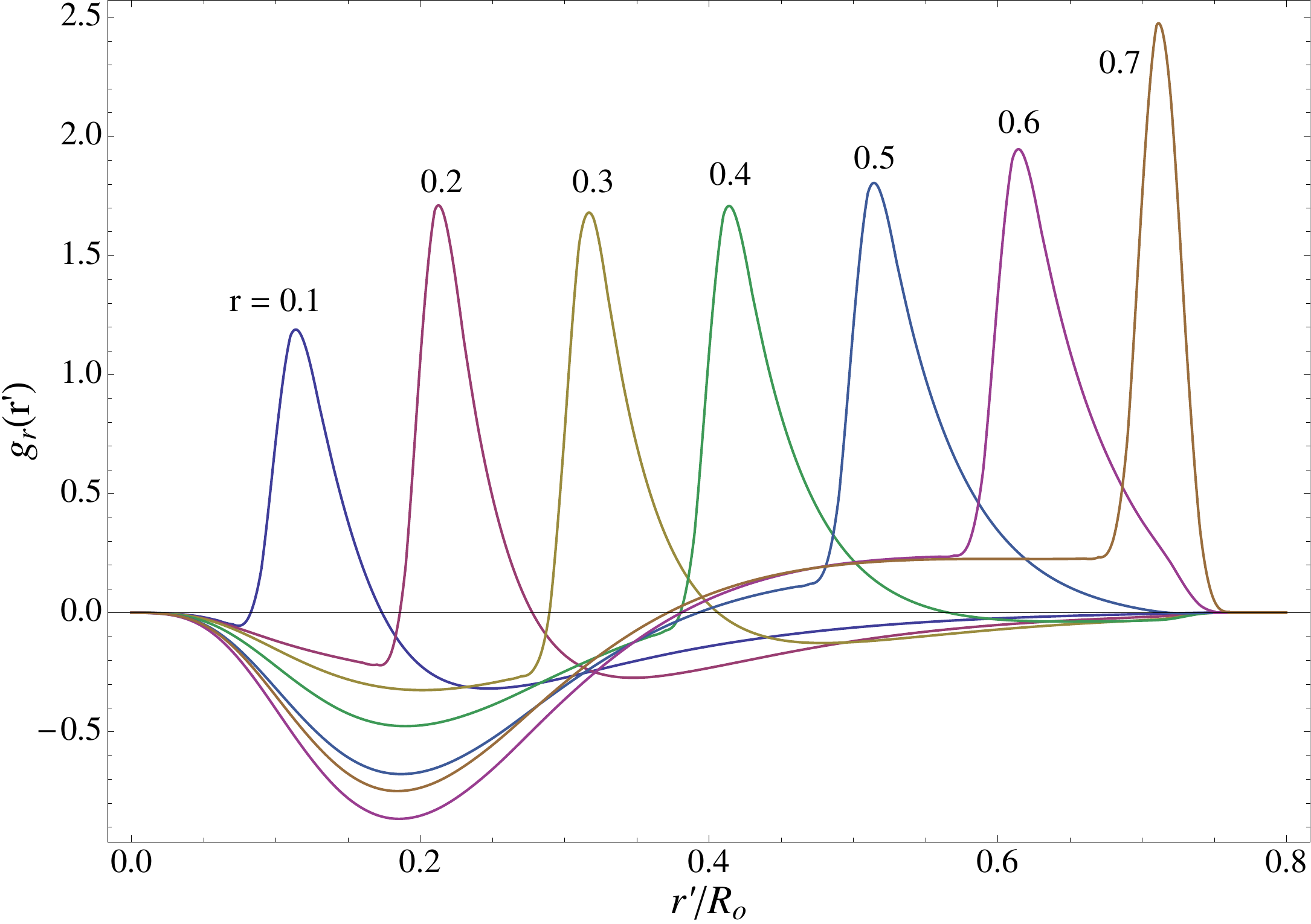}
\end{center}
\par
\vspace{-5mm} \caption{\em {\protect\small Left Panel: the behaviour of 
$f_{r'}(r) \equiv K_u(r,r') R_{\odot}$ as a function of $r$ for the selected values of 
$r'$. 
Right panel: the behaviour of 
$g_{r}(r') \equiv K_u(r,r') R_{\odot}$ as a function of $r'$ for the selected values of 
$r$.  }}
\label{SoundSpeed}
\end{figure}

In Fig.\ref{SoundSpeed}, we discuss the properties of the sound speed kernel $K_{u}(r,r')$ defined by: 
\begin{equation}
\delta u(r) = \int dr' \; K_{u}(r, r') \; \delta \kappa (r')
\end{equation}
where $\delta u(r)$ is the fractional variation of the squared isothermal sound speed $u(r)\equiv P(r)/\rho(r)$.
%
In the left panel of Fig.~\ref{SoundSpeed},  we show the functions 
$f_{r'}(r) \equiv K_u(r,r')\, R_{\odot}$, 
calculated for the selected values $r'= 0.1, 0.2, \dots, 0.7 R_{\odot}$.
In our approach, the functions $f_{r'}(r)$ 
correspond to the sound
speed variations produced (in a linear theory) by  the localised opacity 
increases $G(r-r')$.  One has large effects close to $r'$ which are
due to the variation of the temperature profile of the 
sun, as can be understood by considering that
$\delta u = \delta P - \delta \rho \simeq \delta T -\delta \mu$,
where $\mu$ is the mean molecular weight of the solar plasma. 
From eq.~(\ref{lsm}), we see 
that $\delta T(r)$ is expected to have a sharp decrease  close to $r'$, by an
amount approximately equal to $1/l_t(r')$, while $\delta \mu$ remains approximately constant, 
being related to the chemical composition of the solar plasma.

In the right panel of Fig.~\ref{SoundSpeed} , we show the functions $g_{r}(r')\equiv K_u(r,r')\, R_{\odot}$, 
calculated for the selected values $r= 0.1, \dots, 0.7 R_{\odot}$, 
that quantify the sensitivity of the sound speed 
at a given $r$ to the opacity in the shell $r'$ of the sun. 
They clearly indicates that $u(r)$ is maximally sensitive to the value of the opacity at $r'\simeq r$. 
However, the displayed functions are different from zero everywhere, and have negative values
in large part of the solar radiative region. 
 This suggests that compensating effects can occur, especially when one considers opacity 
modifications that extend over a  broad region of the sun.


In this respect, it is important to note that {\em the sound speed is practically 
insensitive to a global rescaling of opacity}. This can be appreciated by looking at 
the black dotted line in the left panel of  Fig.~\ref{SoundSpeed} , which is defined by:
\begin{equation}
\delta u_{0}(r) = \int dr' K_u(r,r') 
\end{equation}
and corresponds
to the sound speed variation 
produced (in a linear theory) by a constant rescaling $\delta \kappa_{\rm 0}(r) \equiv 1$.
For visualization purposes, it is useful to note that the values of $\delta u_{0}(r)$ at $r= 0.1, \dots, 0.7 R_{\odot}$
correspond to the integral in $r'/R_{\odot}$ of the functions $g_{r}(r')$ 
displayed in the right panel of Fig.~\ref{SoundSpeed}.  We see that $\delta u_{0}(r)$ is very small, 
as a result of an almost perfect compensation between the positive contribution 
from the region $r'\simeq r$ and the negative contribution from the other regions of the sun.

A qualitative argument to explain the stability of the sound speed 
is the following. The virial theorem connects the gravitational energy, $E_{g} = -\int dm \; G m/r$, 
and the thermal energy content of a given star, $E_{i} = 3/2 \int dm \; (P/\rho) = 3/2 \int dm \; u$, 
being $E_g = -2 E_i$. For a generic star,  a 
global rescaling of the opacity 
reflects into a global rescaling  of the radial profile and, thus, taking into account the virial theorem, 
also into a global rescaling of the sound speed.  This is not the case for the sun because 
the solar radius is observationally determined. 
We are forced to re-adjust the free parameters in the model in order to keep the solar radius
fixed, with the effect of stabilizing the radial profile and, thus, the sound speed profile $u(r)$.

\begin{figure}[t]
\par
\begin{center}
\includegraphics[width=8.5cm,angle=0]{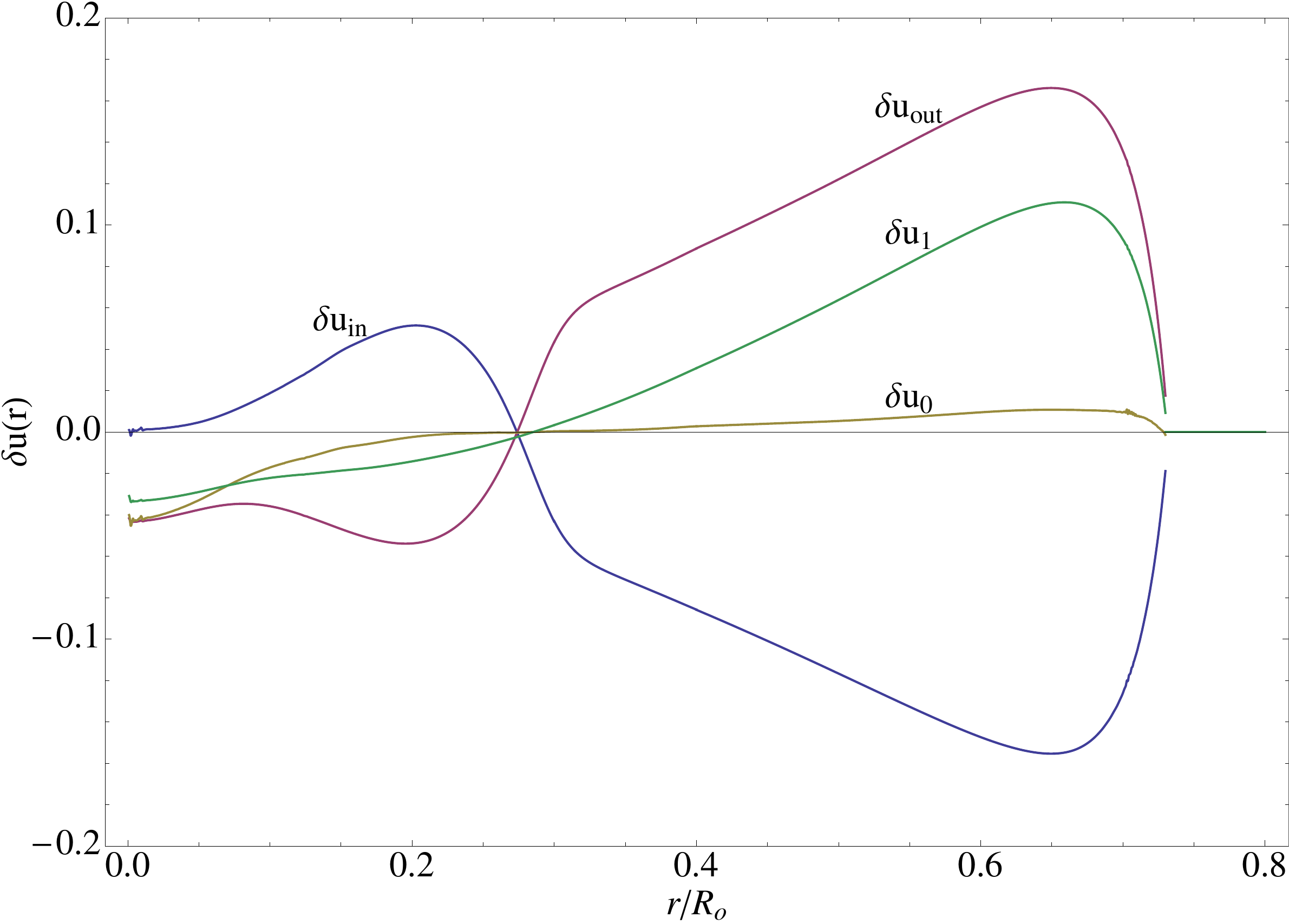}
\includegraphics[width=8.5cm,angle=0]{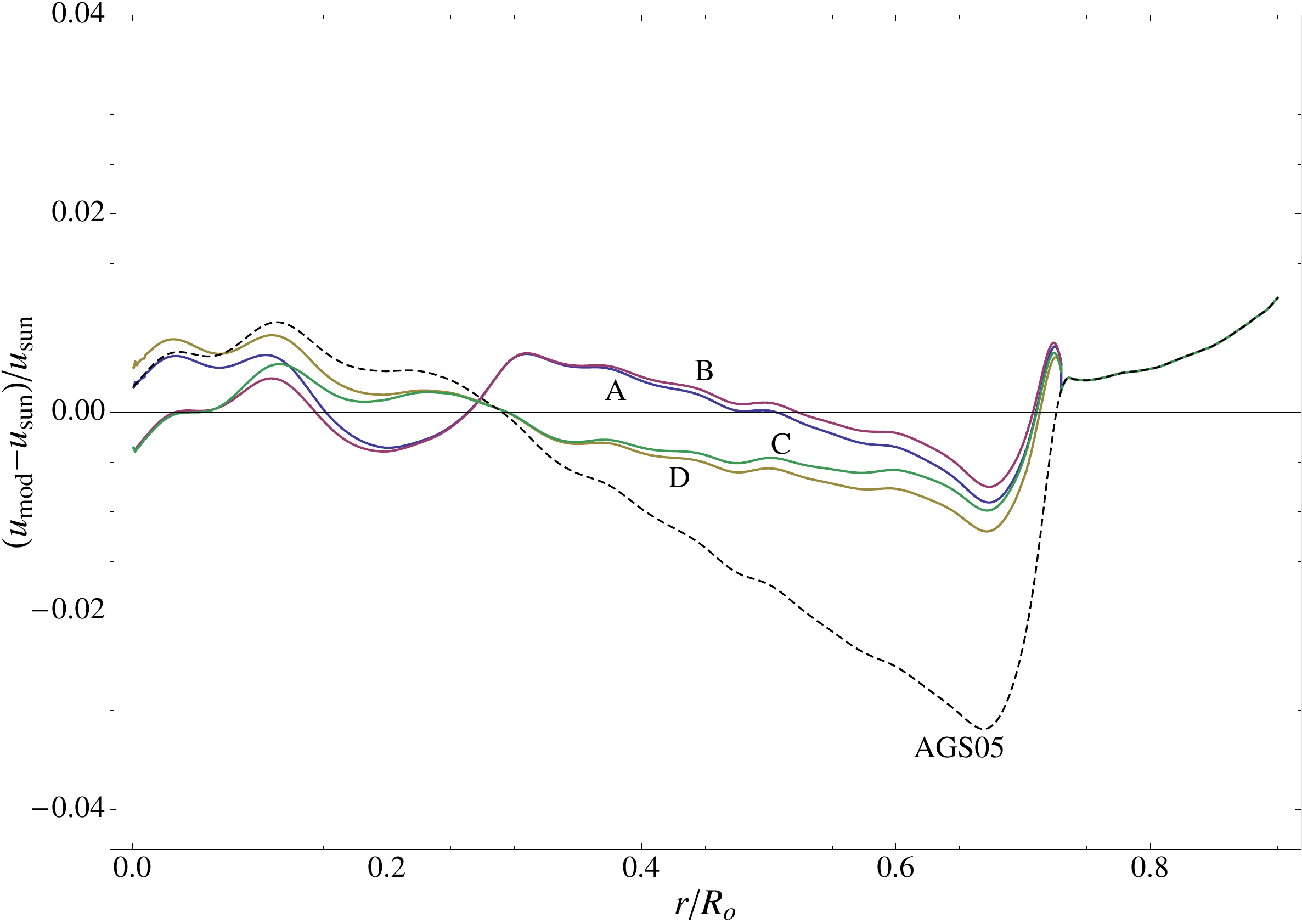}
\end{center}
\par
\vspace{-5mm} \caption{\em {\protect\small Left Panel: The functions $\delta u_{\rm in}(r)$, $\delta u_{\rm out}(r)$, $\delta u_{0}(r)$ and  $\delta u_{1}(r)$
defined in eq.(\ref{u-response}), which allow to calculate the sound speed response to opacity changes parametrized by eqs.(\ref{twozones},\ref{lineartilt}).
Right panel: The difference between the sound speed inferred from helioseismic data and that obtained by solar models implementing AGS05 heavy element 
admixture (dotted line) with suitably modified opacity profile (the solid lines correspond to models A, B, C, D described in the text).}}
\label{SoundSpeed2}
\end{figure}

The above result has relevant implications. In particular, the statement that
 the ``sound speed problem'' requires an increase of the opacity at the bottom of the convective region
is, strictly speaking, not correct because other solutions are possible. 
We can consider, e.g, the sound speed profiles produced by
the opacity variations parametrized by eqs.~(\ref{twozones},\ref{lineartilt}). 
In linear approximation, we have that:
\begin{eqnarray}
\nonumber
\delta u(r) &=& A_{\rm in} \, \delta u_{\rm in}(r) + A_{\rm out} \, \delta u_{\rm out}(r)\\
\label{u-response}
\delta u(r) &=& A_{0} \, \delta u_{0}(r) + A_{1} \, \delta u_{1}(r) \, .
\end{eqnarray}
The functions $\delta u_{\rm in}(r)$, $\delta u_{\rm out}(r)$,
$\delta u_{0}(r)$ and $\delta u_{1}(r)$ are shown in the left panel of Fig.~\ref{SoundSpeed2} 
and are reported in Tab.~\ref{tab2} for the selected values $r=0.1, 0.2, 0.4, 0.65 \, R_{\odot}$.
We see that $\delta u_{0}(r) \ll \delta u_{1}(r)$, indicating that the  ``tilt'' and not the scale of the opacity is
fixed by the sound speed. Moreover, we have $\delta u_{\rm in}(r)\simeq - \delta u_{\rm out}(r)$
that shows that the effect of an enhancement of the opacity in the external radiative region can be
equally produced by a decrease of the opacity at the solar center.


This is confirmed in the right panel of Fig.\ref{SoundSpeed2} where we show the sound speed profiles obtained
by implementing four different opacity modifications, described by:
\par\noindent
{\em Model A:} 15\% decrease of opacity in the energy producing region 
($A_{\rm in} =- 0.15$ and $A_{\rm out} = 0$);
\par\noindent
{\em Model B:} 15\% increase of the opacity in the
outer radiative region 
($A_{\rm in} =0$ and $A_{\rm out} = 0.15$);
\par\noindent
{\em Model C:} a linear tilt of opacity
corresponding to a 20\% increase at the bottom of the convective envelope ($A_{0} = 0$ and $A_{1}=0.2$);
\par\noindent
{\em Model D:} a linear tilt plus a global rescaling of opacity 
corresponding to a 20\% decrease at the solar center ($A_{0} = -0.2$ and $A_{1}=0.2$).
\par\noindent
The sound speed profiles obtained in all the considered cases
reproduce equally well the result inferred by helioseismic data, showing that
the helioseismic determination of $u(r)$ translates into a bound on the differential increase
$A_{\rm out}- A_{\rm in}  \simeq 0.15$ and on the tilt $A_{1}\simeq 0.2$, with no relevant constraint on $A_{0}$ 
(or, equivalently, $A_{\rm in}+A_{\rm out}$) that fix the global scale of opacity.

In light of this observation, it is intriguing the possibility that non-standard effects that mimic a decrease
of the opacity at the solar center, like e.g. the accumulation of few GeVs WIMPs in 
the solar core (see e.g. \cite{WimpsNoi}), could have a role in the solution of the solar composition puzzle.
We will see, in the following, that this possibility is disfavoured by the determination
of the surface helium abundance and by the measurement of the $^{7}{\rm Be}$ and $^{8}$B neutrino fluxes\footnote{The 
idea that the accumulation of few GeV WIMPs in the solar core could alleviate the 
``solar composition problem'' was originally proposed by \cite{Taup}. The recent paper \cite{WimpsSarkar} 
presented a qualitative implementation of this idea. The effect on boron neutrinos is discussed in \cite{bertone}.}.

\subsection{The  surface helium abundance}

\begin{figure}[t]
\par
\begin{center}
\includegraphics[width=8.5cm,angle=0]{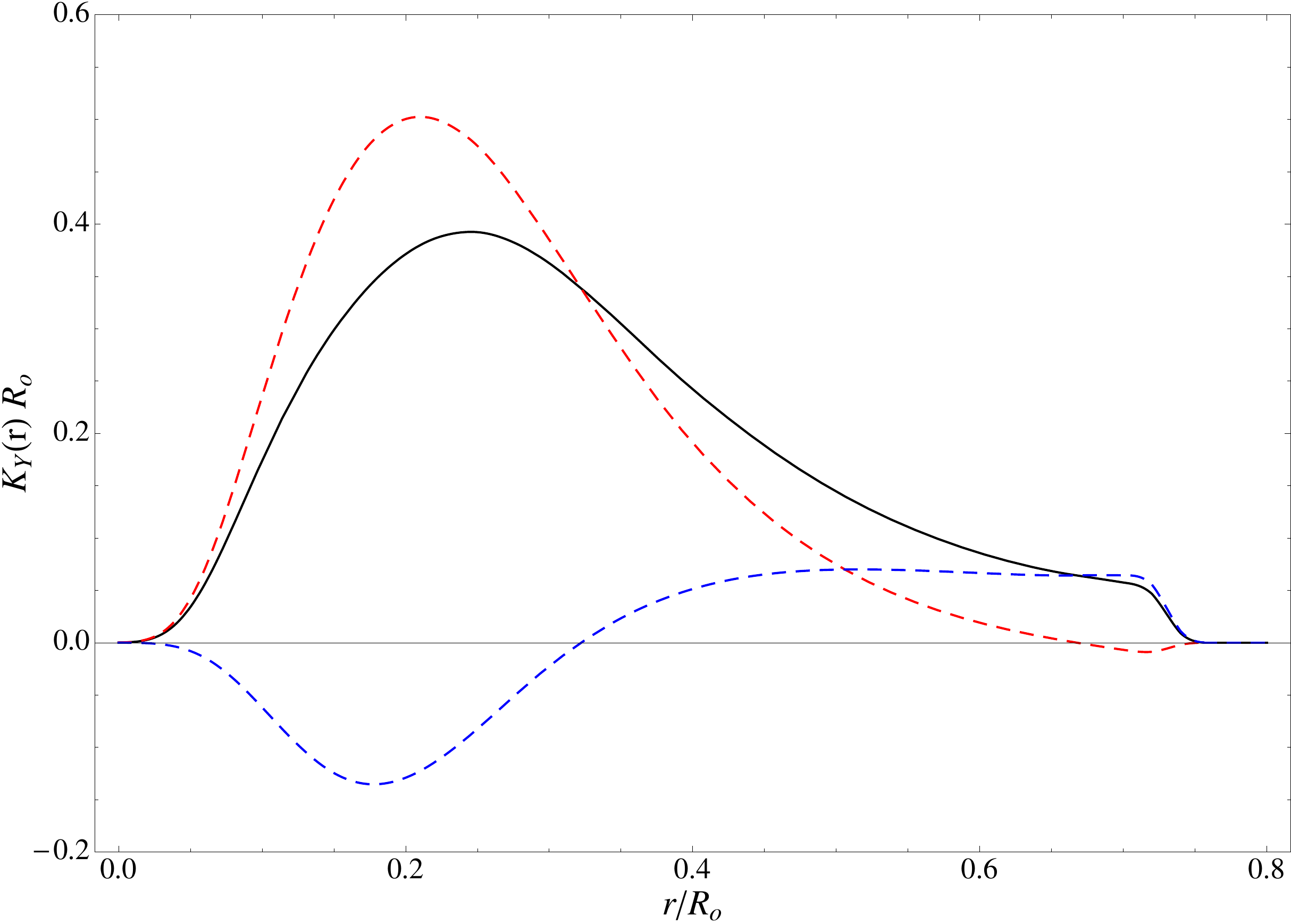}
\includegraphics[width=8.5cm,angle=0]{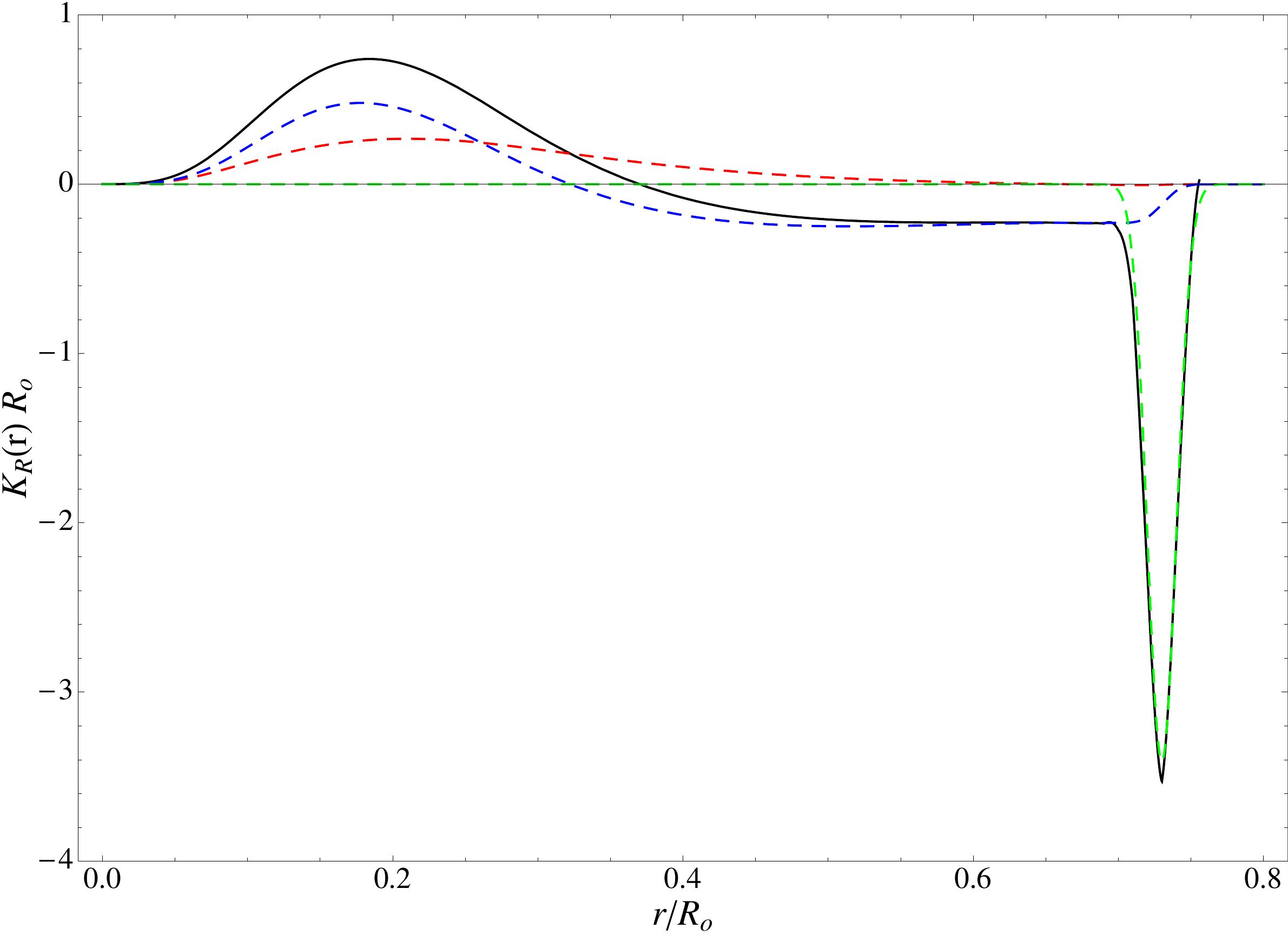}
\end{center}
\par
\vspace{-5mm} \caption{\em {\protect\small  Left Panel: The solid line corresponds to the kernel $K_{Y}(r)$ defined in eq.(\ref{Helium}). 
The red and blue dashed lines describe the contributions to surface helium variations provided by
the terms $A_{Y}\Delta Y_{\rm ini}$ and $A_{C} \, \delta C$, respectively. See text for details.
Right Panel: The solid line corresponds to the kernel $K_{R}(r)$ defined in eq.(\ref{Radius}). 
The red, blue and green dashed lines describe the contributions to convective radius variations provided by
the terms $\Gamma_{Y}\Delta Y_{\rm ini}$, $\Gamma_{C} \, \delta C$ and $\Gamma_{\kappa} \,\delta\kappa_{\rm b}$, 
respectively. See text for details.}}
\label{HeliumandRadius}
\end{figure}

In the left panel of Fig.\ref{HeliumandRadius}, we show with solid line the functional derivative $K_{Y}(r)$ of the 
surface helium abundance $Y_{\rm b}$, defined according to equation:
\begin{equation}
\Delta Y_{\rm b} = \int dr\; K_{\rm Y}(r) \; \delta \kappa(r) 
\label{Helium}
\end{equation}
where we considered the {\em absolute} variation $\Delta Y_{\rm b}$ to conform 
with the notations adopted in \cite{noi}\footnote{Here and in the following, we use the notation $Q_{\rm b}$ to indicate 
that a given quantity $Q(r)$ is evaluated at the bottom of the convective region, i.e. $Q_{\rm b} \equiv Q(\overline{R}_{\rm b})$ 
where $\overline{R}_{\rm b}=0.730 \, R_{\odot}$.}.

 The surface helium abundance depends on the initial chemical composition and on the effects of elemental diffusion. 
In ref.\cite{noi}, by assuming that (the variation of) the time-integrated effect
of diffusion 
can be estimated from (the variation of) diffusion efficiency in the present sun, 
we have obtained the following relation:
\begin{equation}
\Delta Y_{\rm b} = A_{Y} \; \Delta Y_{\rm ini} + A_{C} \; \delta C
\label{HeliumLSM}
\end{equation} 
with $A_{Y}=0.838$ and $A_{C}=0.033$, which gives $\Delta Y_{\rm b}$ as a function of  the absolute variation of the initial helium abundance, $\Delta Y_{\rm ini}$, 
and of the fractional variation of pressure at the bottom of the convective region, $\delta C= \delta P_{\rm b}$. 
The above equation allows us to obtain the functional derivative $K_{Y}(r)$ as the sum of two contributions, 
one related to the term $A_{Y}\; \Delta Y_{\rm ini}$ and the other to 
$A_{\rm C} \; \delta C$. These
are shown in Fig.\ref{HeliumandRadius} with red and blue dashed lines, respectively.

The function $K_{\rm Y}(r)$ is positive everywhere, showing that an increase of 
opacity in an arbitrary shell of the sun translates into an increase of the helium abundance\footnote{
If the chemical composition is fixed, an increase of the opacity 
implies a decrease of the total luminosity $L$ which roughly 
scales as $L\propto \mu^4/\kappa $, where $\mu$ is the mean molecular weight. 
In order to reproduce the observed solar luminosity $L_{\odot}$, 
we are forced to readjust the chemical composition of the sun by increasing the helium abundance. 
This has the simultaneous effects of  increasing $\mu$ and decreasing $\kappa$.}.
Moreover, the kernel $K_{Y}(r)$ has a rather broad profile. 
This implies that the determination of $Y_{\rm b}$ effectively constrains the 
opacity scale and breaks the degeneracy between the 
possible solutions of the sound speed problem presented in the previous section.
The SSM that implements the AGS05 heavy element admixture predicts  
the value $\overline{Y}_{\rm b}=0.229 $ (see e.g. \cite{noi}) 
which is about 6$\sigma$ lower 
than the helioseismic determination $Y_{\rm b} =0.2485 \pm 0.0034$ \cite{basu}.
This discrepancy requires an increase of the opacity, as it can be obtained, e.g., 
by increasing the metal content of the sun. 
Models that accounts for a reduction 
``effective'' opacity at the solar center are expected
to decrease $Y_{\rm b}$, increasing the disagreement with helioseismic results.


A simple quantitative analysis can be performed by considering the opacity profiles (\ref{twozones},\ref{lineartilt})
that produce the variations $\Delta Y_{\rm b}$ given by:
\begin{eqnarray}
\label{twozonesHelium}
\Delta Y_{\rm b} &=&  0.073 \; A_{\rm in} + 0.069 \; A_{\rm out} \simeq 0.07 \; ( A_{\rm in}  +  A_{\rm out})\\
\label{lineartiltHelium}
\Delta Y_{\rm b}  &=& 0.142 \; A_{0} + 0.062 \; A_{1}
\end{eqnarray}
By comparing the SSM prediction with the helioseismic result, we obtain $A_{\rm in} + A_{\rm out} = 0.28 \pm 0.05$ that, 
combined with the ``orthogonal'' constraint $A_{\rm out} - A_{\rm in} \simeq 0.15$ provided by the sound 
speed, gives  $A_{\rm in}= 0.04 \div 0.09$ and $A_{\rm out}= 0.19 \div 0.24$. 
Alternatively, we can use eq.(\ref{lineartiltHelium}) and the information on the tilt $A_1\simeq 0.2$ provided by the
sound speed, to fix the opacity at the solar center obtaining $A_{0}=0.035 \div 0.075$.


\subsection{The  convective radius}

 In the right panel of Fig.\ref{HeliumandRadius},  we show with a solid line the functional derivative $K_{R}(r)$ 
of the convective radius $R_{\rm b}$ defined according to the relation:
\begin{equation}
\delta R_{\rm b} = \int dr\; K_{\rm R}(r) \; \delta \kappa(r) 
\label{Radius}
\end{equation} 
We see that the kernel $K_{R}(r)$ has a very sharp peak at  $r \simeq \overline{R}_{\rm b}= 0.730 R_{\odot}$ 
that reflects the (well-known) fact that the convective radius is particularly sensitive to the opacity at the
bottom of the convective region. 
The shape of this peak has not a precise physical meaning and depends on the method of calculation.
The effect of opacity changes has been, in fact, estimated within the LSM approach, in which the 
fractional variation $\delta R_{\rm b}$ is calculated from:
\begin{equation}
\delta R_{\rm b} = \Gamma_{Y} \; \Delta Y_{\rm ini} + \Gamma_{C} \; \delta C +  \Gamma_\kappa \; \delta \kappa_{\rm b}
\label{RadiusLSM}
\end{equation}
where $\Gamma_{Y}=0.449$, $\Gamma_{C}=-0.117$ and $\Gamma_\kappa = -0.085$, while $\delta \kappa_{\rm b}$
is fractional variation of opacity at the bottom of the convective envelope,
i.e. $\delta \kappa_{\rm b} = \delta \kappa(\overline{R}_{\rm b})$.
The ``local'' term $\Gamma_\kappa \, \delta \kappa_{\rm b}$ 
translates into a delta-function contribution
$\Gamma_\kappa \, \delta(r-\overline{R}_{\rm b})$ to the 
functional derivative. 
Since we evaluate numerically the kernel by applying
a localised gaussian increase of opacity, this is
convolved with the function 
$G(r-r_0)$ given in eq.(\ref{gaussianincrease}).
As a final result, one obtains the contribution $\Gamma_\kappa \, G(r-\overline{R}_{\rm b})$ to the kernel $K_{R}(r)$
which is shown by the green dashed line in Fig.\ref{HeliumandRadius}, whereas the red and blue dashed lines describe the contributions
arising from $\Gamma_Y\; \Delta Y_{\rm ini}$ and $\Gamma_{C} \;\delta C$, respectively.
We remark that the area under the peak at $r=\overline{R}_{\rm b}$ is approximately equal to $\Gamma_{\kappa}$ and 
does not depend on the calculation method. We can, thus, safely use the functional derivative $K_{R}(r)$
to describe all the situations in which opacity varies on scale larger than $\delta r = 0.01 R_{\odot}$. 

Eqs.(\ref{HeliumLSM}) and (\ref{RadiusLSM}) can be combined to obtain a direct determination of $\delta \kappa_{\rm b}$ from
quantities that are all determined by helioseismic observations. We can, in fact, eliminate $\Delta Y_{\rm ini}$ from 
eq.(\ref{RadiusLSM}), obtaining:
\begin{equation}
\delta \kappa_{\rm b} = C_{Y}\;\Delta Y_{\rm b} + C_{\rm R} \; \delta R_{\rm b} + C_{\rho} \; \delta \rho_{\rm b} 
\label{dkappab}
\end{equation}
where:
\begin{eqnarray}
C_{Y} &=& - \frac{\Gamma_{Y}}{A_{Y}\;\Gamma_{\kappa}} = 6.27\\
C_{\rm R} &=& \frac{1}{\Gamma_{\kappa}} = -11.71 \\
C_{\rho} &=&  \frac{1}{\Gamma_{\kappa}}\left[\frac{A_{C}\; \Gamma_{Y}}{A_{Y}}-\Gamma_{C}\right] = -1.58
\end{eqnarray}
In the derivation of the above result, we have considered that $\delta C = \delta P_{\rm b} \simeq \delta \rho_{\rm b}$,
since the fractional variation of the sound speed $\delta u(r)$ is expected to vanish at the bottom of the convective region, 
i.e. $\delta u_{\rm b} = \delta P_{\rm b} - \delta \rho_{\rm b}\simeq 0$, as it is discussed in \cite{noi}.
The discrepancy between the helioseismic determinations of $R_{\rm b}$ and $Y_{\rm b}$ and 
the predictions of SSMs implementing AS05 admixture is quantified as 
$\delta R_{\rm b}=-0.0205 \pm 0.0015$ and  $\Delta Y_{\rm b}=0.0195 \pm 0.0034$. 
The density at the bottom of the convective region deviates from 
value inferred by helioseismology by $\delta \rho_{\rm b} = 0.08 $, as it is discussed e.g. in \cite{serenelli}.
We obtain $\delta \kappa_{\rm b} \simeq 0.24 \pm 0.03$, where errors have been combined in quadrature and
we have neglected the (sub-dominant) contribution to the total error budget due to the uncertainties in the
density determination. We remark that the obtained results is model-independent, since it does not rely on 
any assumption or parametrization for the function $\delta \kappa(r)$.

As a final application, we consider the response of $R_{\rm b}$ to the 
opacity changes parametrized by eqs.(\ref{twozones},\ref{lineartilt}).  We obtain the relations
\begin{eqnarray}
\delta R_{\rm b} &=& 0.12 \;  A_{\rm in} - 0.14 \; A_{\rm out} \simeq 0.13 \; (A_{\rm in} - \; A_{\rm out}  ) \\
\delta R_{\rm b} &=& -0.02 \; A_{0} -0.10  \; A_{1}
\end{eqnarray}
which show that the convective radius, just like the sound speed, provides bounds 
on the differential increase $A_{\rm out}- A_{\rm in}$ and on tilt $A_{1}$, 
with no relevant constraints on $A_{0}$ (or equivalently $A_{\rm in} + A_{\rm out}$) that fix the
global scale of opacity. 
By considering $\delta R_{\rm b} = -0.0205 \pm 0.0015$, we obtain
$A_{\rm out}-A_{\rm in} \sim 0.15 $ and $A_{1} \sim 0.2$ 
in substantial agreement (and complete degeneracy) with the information provided by the sound 
speed measurement. By combining this information with the 
constraints provided by the surface helium abundance and performing a simple $\chi^2$ analysis, 
we obtain $A_{\rm in} = 0.07 \pm 0.04$ and $A_{\rm out} = 0.21 \pm 0.04$ for the parametrization 
given in eq.~(\ref{twozones}) and $A_{0} = 0.056 \pm 0.040$ and $A_{1} = 0.187 \pm 0.023$ for that given in eq.~(\ref{lineartilt}).
The corresponding bounds on the opacity change $\delta \kappa(r)$ are shown in fig.~(\ref{FigFinal}) and commented 
in the conclusive section.


\begin{figure}[t]
\par
\begin{center}
\includegraphics[width=8.5cm,angle=0]{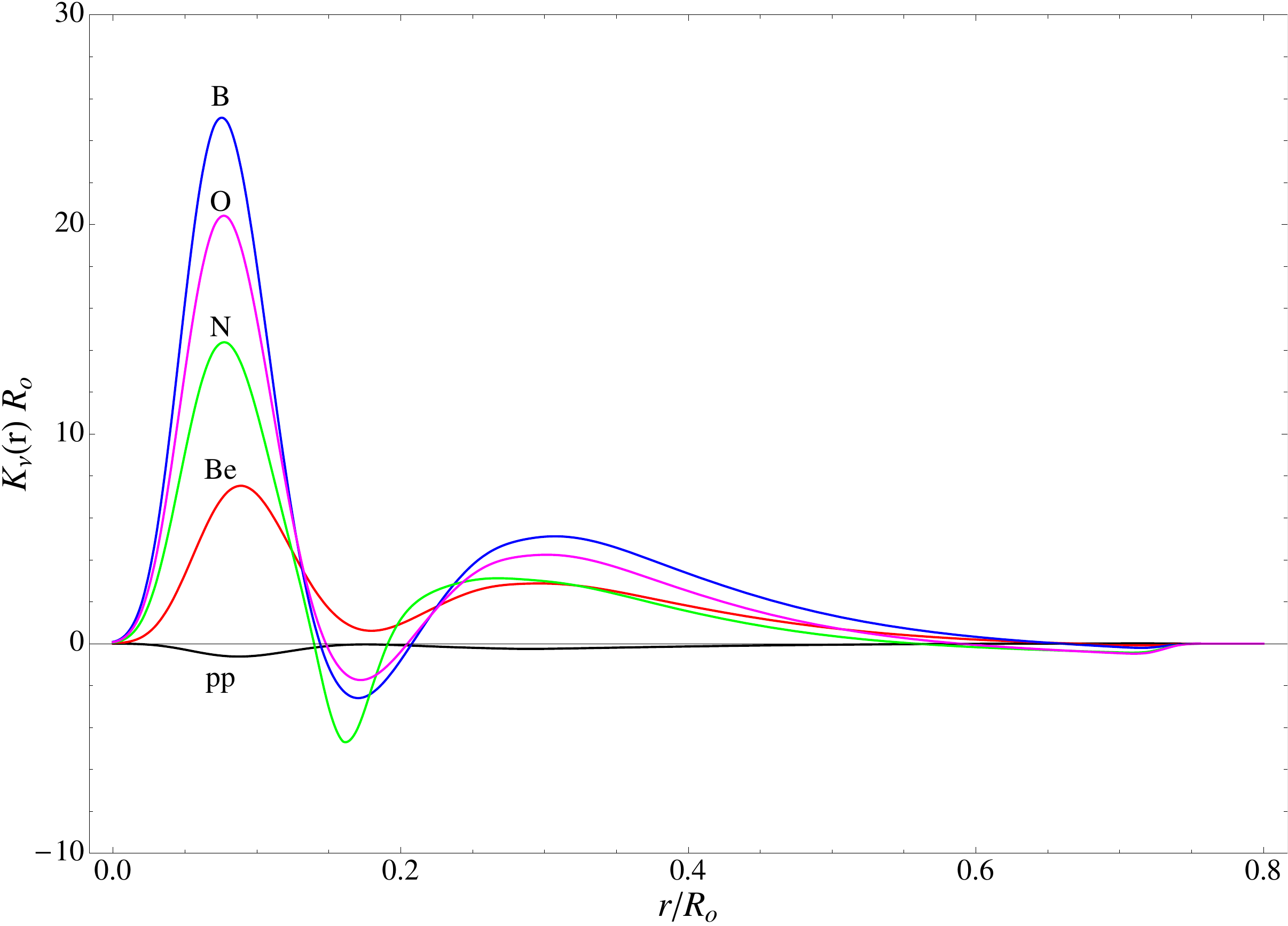}
\includegraphics[width=8.5cm,angle=0]{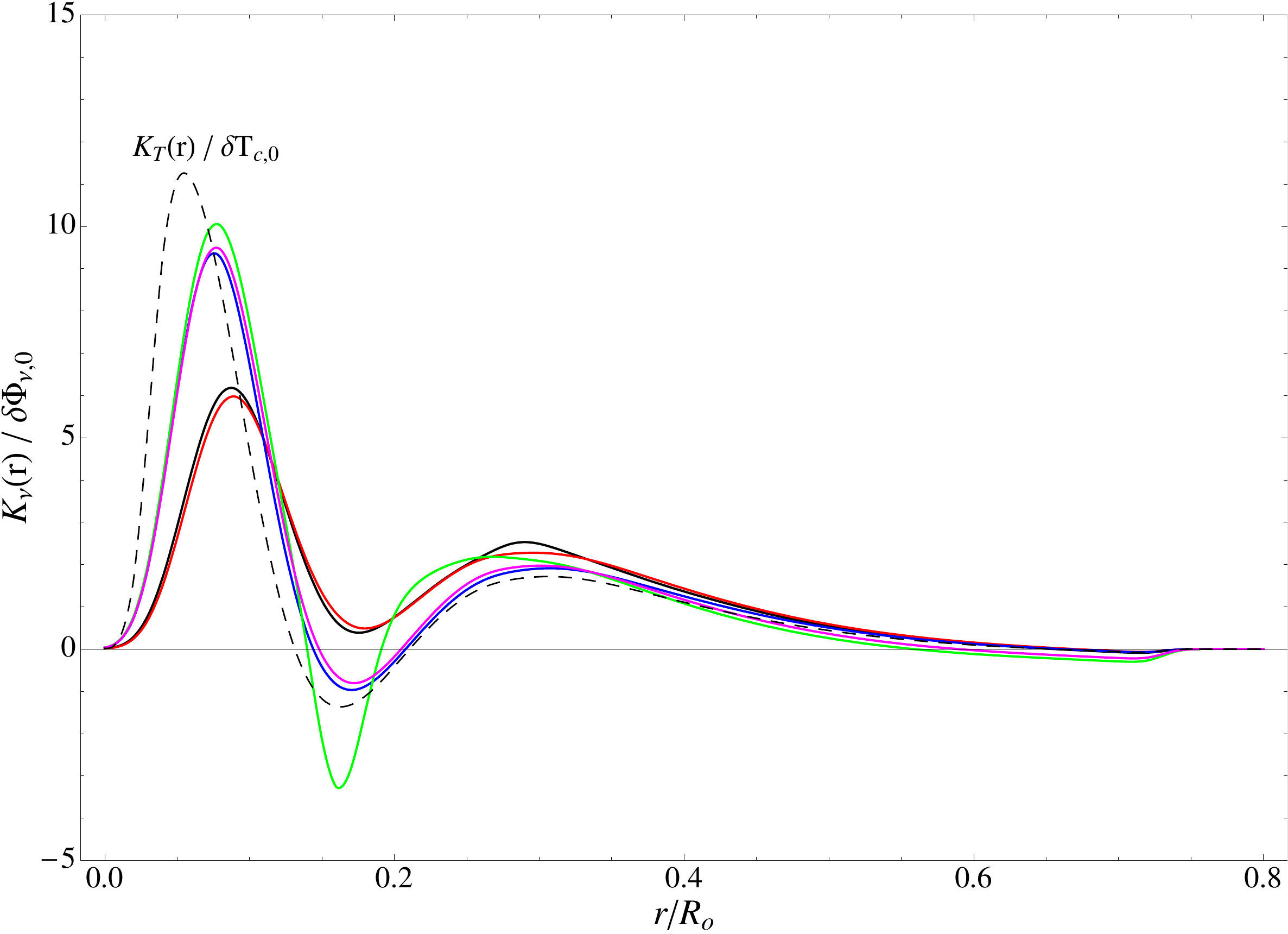}
\end{center}
\par
\vspace{-5mm} \caption{\em {\protect\small  Left Panel: The solar neutrino kernels $K_\nu(r)$ defined in eq.(\ref{nu-kernels}).
Right Panel: The solid lines are the normalized solar neutrino kernels $K_\nu(r)/ \delta \Phi_{\nu,0}$. The dashed line shows the normalized 
kernel $K_{T}(r)/\delta T_{{\rm c},0}$ defined in eq.(\ref{T-kernel}), 
that describes the response of the solar central temperature to localised opacity modifications.}}
\label{Figneutrinos}
\end{figure}


\subsection{Neutrino Fluxes}

In the left panel of Fig.\ref{Figneutrinos},  we show the functional derivatives $K_{\nu}(r)$
of the neutrino fluxes $\Phi_\nu$ defined according to relation:
\begin{equation}
\delta \Phi_{\nu} = \int dr\; K_{\nu}(r) \; \delta \kappa(r) 
\label{nu-kernels}
\end{equation} 
where the index $\nu ={\rm pp},\, {\rm Be},\,{\rm B}, \, {\rm N}, \, {\rm O}$ labels the neutrino producing reactions according to the usual convention.
The kernel $K_{\nu}(r)$ have been calculated by using the LSM approach 
and by taking into account that the fractional variations of the fluxes $\delta \Phi_{\nu}$ 
are related to physical and chemical properties of the sun by:
\begin{equation}
\delta \Phi_\nu = \int dr \left[ \phi_{\nu,\rho}(r) \, \delta \rho(r) +\phi_{\nu,T}(r) \, \delta T(r)+ \phi_{\nu,Y}(r) \, \Delta Y(r) + \phi_{\nu,Z}(r) \, \delta Z(r) \right]
\end{equation}
The functions $\phi_{\nu,j}(r)$ have been defined and calculated in the fig.~10 of \cite{noi}.

  Our results show that neutrino fluxes probe the opacity of the sun in the region $r\le 0.45 R_{\odot}$. 
The kernels $K_{\rm B}(x)$, $K_{\rm Be}(x)$, $K_{\rm N}(x)$ and $K_{\rm O}(x)$ 
are positive-valued almost everywhere, while the kernel $K_{\rm pp}(r)$ is negative. 
This indicates that an increase of opacity 
generally translates into an enhancement of $^{8}{\rm B}$, $^{7}{\rm Be}$ and CNO neutrino fluxes 
and into a (slight) decrease of the ${\rm pp}-$neutrino component.
In Tab.~\ref{tab2}, we show the coefficients $\delta \Phi_{\nu,\rm in}$, $\delta \Phi_{\nu,\rm out}$, $\delta \Phi_{\nu, 0}$
and $\delta \Phi_{\nu, 1}$ that allows to describe the effects of opacity changes parametrized by 
eqs.~(\ref{twozones},\ref{lineartilt}), through the simple relations:
\begin{eqnarray}
\delta \Phi_\nu &=& A_{\rm in} \; \delta \Phi_{\nu,\rm in} + A_{\rm out} \;\delta \Phi_{\nu,\rm out}\\
\delta \Phi_\nu &=& A_{0} \; \delta \Phi_{\nu,0} + A_{1} \;\delta \Phi_{\nu,1} ~.
\end{eqnarray}
We see that $^8$B and CNO neutrinos are extremely sensitive to opacity changes,
as it is expected since they strongly depend on the temperature of the central regions of the sun.
It is interesting to note in the right panel of Fig.~\ref{Figneutrinos} that the normalized neutrino kernels
$K_{\nu}(x) /\Phi_{\nu,0 }$  have a common behaviour,
with two maxima at $r \sim 0.1 R_{\odot}$ and $r\sim 0.3 R_{\odot}$,  
and one minimum at $r\sim 0.2 R_{\odot}$.  This shows that different fluxes 
basically probe the same quantity, constraining the  opacity profile 
 with the maximal sensitivity in the two regions of the sun $r \sim 0.05 \div 0.15$  and $r \sim 0.2 \div 0.45$.
 
Some insights on the above results can be obtained by recalling that 
the total neutrino flux is essentially fixed by the solar luminosity constraint. We, thus, expect that
$\Delta \Phi_{\rm tot} = \sum_{\nu} \overline{\Phi}_{\nu}  \cdot \delta \Phi_{\nu} \simeq 0$, 
where $\overline{\Phi}_{\nu}$ are the SSM predictions for the various neutrino components.
Considering that about 99\% of the total flux is provided by pp and Be neutrinos,
the luminosity constraint implies 
$ \delta \Phi_{\rm pp} \simeq -(\overline{\Phi}_{\rm Be} /\overline{\Phi}_{\rm pp} ) \, \delta \Phi_{\rm Be} = -0.075\,  \delta \Phi_{\rm Be}$ 
which explains the smallness of the pp-neutrino kernel, the ratio between the pp and Be-neutrino coefficients 
in tab.~\ref{tab2} and the equality between the normalized pp and Be-neutrino kernels observed in the right 
panel of Fig.~\ref{Figneutrinos}.
 The wavy shape of the kernels $K_\nu(r)$ reflects, instead, the response 
of central temperature to localised opacity modifications. This was first noted and discussed 
by \cite{chris} and it is seen in the right panel of Fig.~\ref{Figneutrinos}, 
where we show with a dashed line the functional derivative of the central temperature $T_{\rm c}$, 
defined by:
\begin{equation}
\delta T_{\rm c} = \int dr  \; K_{T}(r) \; \delta \kappa(r)
\label{T-kernel}
\end{equation}
For convenience, we plot the normalized kernel $K_{\rm T}(x)/ \delta T_{{\rm c}, 0} $, where the normalization
factor is $\delta T_{{\rm c}, 0} = 0.138$. 

\begin{figure}[t]
\par
\begin{center}
\includegraphics[width=8.5cm,angle=0]{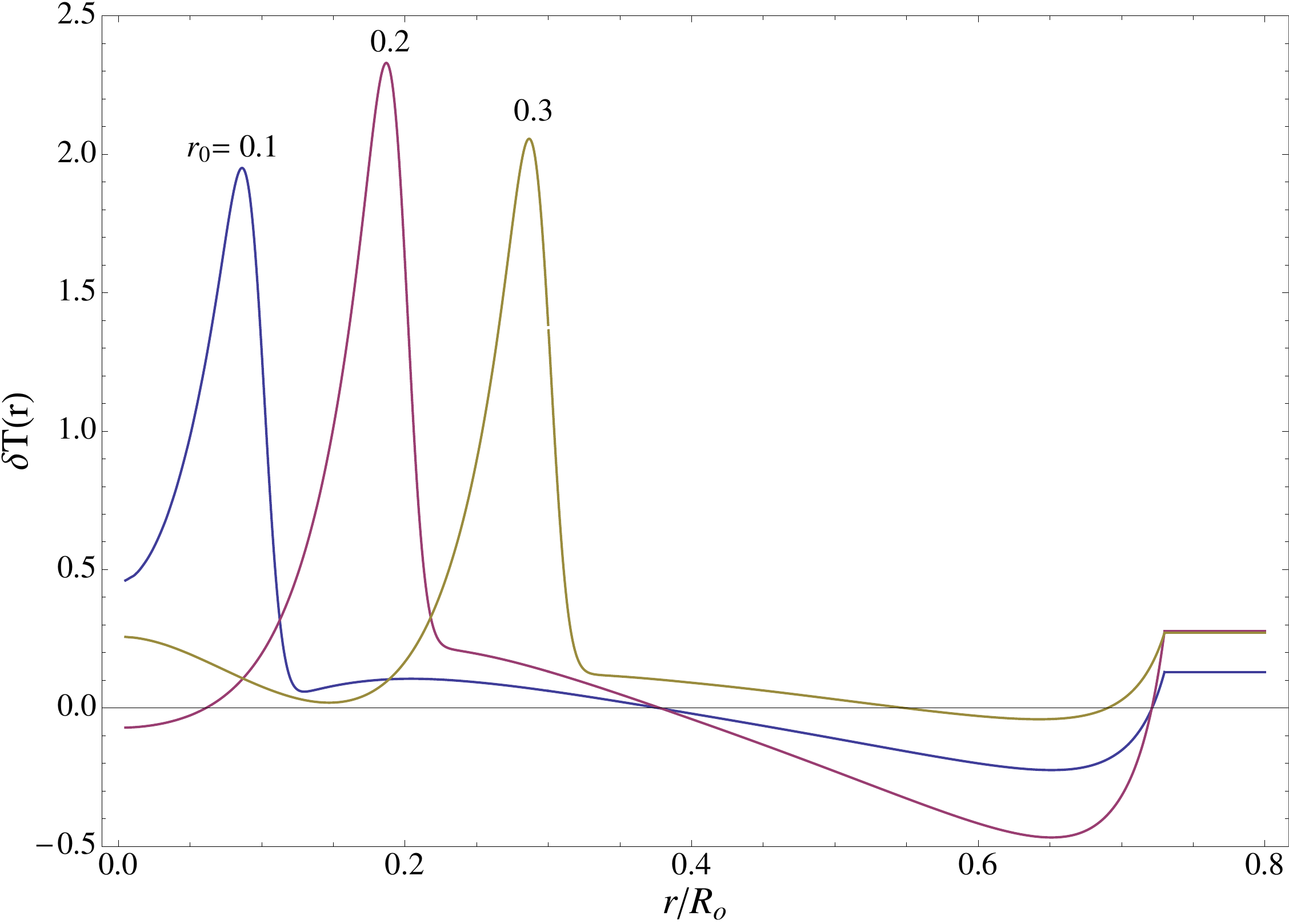}
\end{center}
\par
\vspace{-5mm} \caption{\em {\protect\small  The temperature profile variations $\delta T(r)$ produced (in LSM) by the localized 
gaussian increases of opacity $G(r-r_{0})$ defined in eq.(\ref{gaussianincrease}) with $r_0 = 0.1, \, 0.2,\, 0.3 \,R_{\odot}$.}}
\label{figT}
\end{figure}

The peculiar behaviour with $r$ of the kernel $K_{T}(r)$
cannot be explained in simple terms. A qualitative comprehension 
can be obtained from Fig.~\ref{figT}, where we show the temperature profiles $\delta T(r)$
produced by the opacity changes $\delta \kappa(r)=G(r-r_0)$ with
$r_0 = 0.1,\, 0.2,\, 0.3 \; R_{\odot}$. We see that the performed opacity modifications
translate into a large increase of temperature close to $r_{0}$,
that necessarily alters nuclear burning rates since $r_{0}$ is inside or close to the energy producing region.
The free parameters of the solar model are readjusted
in such a way that the same luminosity is obtained with a different temperature profile. 

In the two cases $r_{0}=0.1 R_{\odot}$ (i.e. opacity increase well inside the energy producing region)
and $r_{0}=0.3 R_{\odot}$ (i.e. opacity increase just outside the energy producing region), 
the entire energy producing zone is affected. A new ``equilibrium'' situation 
is achieved in which nuclear burnings occur at higher temperatures with a larger helium abundance,
favoring $^8$B, CNO and $^7$Be neutrinos at expenses of pp-neutrinos.
For $r_{0}=0.2 R_{\odot}$, we have a peculiar situation since the maximal effect
is produced in a region of the sun where only pp neutrinos are produced, as it is seen from 
fig.10 of \cite{noi}. 
This necessarily shifts energy production outwards with respect to the SSM case. 
In order to avoid overproduction of energy and to recover the observed luminosity,
the central temperature $T_{\rm c}$ has to be (slightly) decreased so that the burning rates 
at the center of the sun are suppressed.


At present, the best studied components of the solar neutrino flux are the $^8{\rm B}$ neutrino 
flux which is determined by the SNO neutral current measurement with about 6\% accuracy,
$\Phi_{\rm B} =(5.18 \pm 0.29) \times 10^6 \;{\rm cm}^{-2}\; {\rm s}^{-1}$ \cite{SNO},
and the $^7$Be neutrino flux which is measured by Borexino,
$\Phi_{\rm Be} =(5.18 \pm 0.51) \times 10^9 \;{\rm cm}^{-2}\; {\rm s}^{-1}$ \cite{Borex}.
These fluxes have to be compared with the results of theoretical calculations. 
SSMs implementing AGS05 heavy elements admixture predict values for $\Phi_{\rm B}$ and $\Phi_{\rm Be}$
which are about 10\% lower than the experimental results. 
We take, for definiteness, the values $\overline{\Phi}_{\rm B} = 4.66 \times 10^6 \;{\rm cm}^{-2}\; {\rm s}^{-1}$
and  $\overline{\Phi}_{\rm Be} = 4.54 \times 10^9 \;{\rm cm}^{-2}\; {\rm s}^{-1}$
obtained in \cite{serenelli} which are affected by $\sim 9\%$ and $\sim 5\%$ 
theoretical uncertainties\footnote{The quoted theoretical uncertainties are obtained from \cite{penagaray}. In our estimate, 
we have not included the contribution due to opacity, since this is considered as a free parameter in our analysis.}, respectively.
The difference between the theoretical predictions 
and the experimental data points toward a moderate increase of the central opacity of the sun.
As an example, a 5\% increase of the opacity in the region $r \le 0.3 R_{\odot}$
would produce a 9.7\% (4.3\%) increase of the $^8 {\rm B}$ ($^7{\rm Be}$) neutrino flux. 
Models that account for a reduction of the ``effective'' central opacity, due e.g. to WIMP accumulation, increase
the disagreement and are, thus, disfavoured by solar neutrino data. 

\begin{figure}[t]
\par
\begin{center}
\includegraphics[width=8.5cm,angle=0]{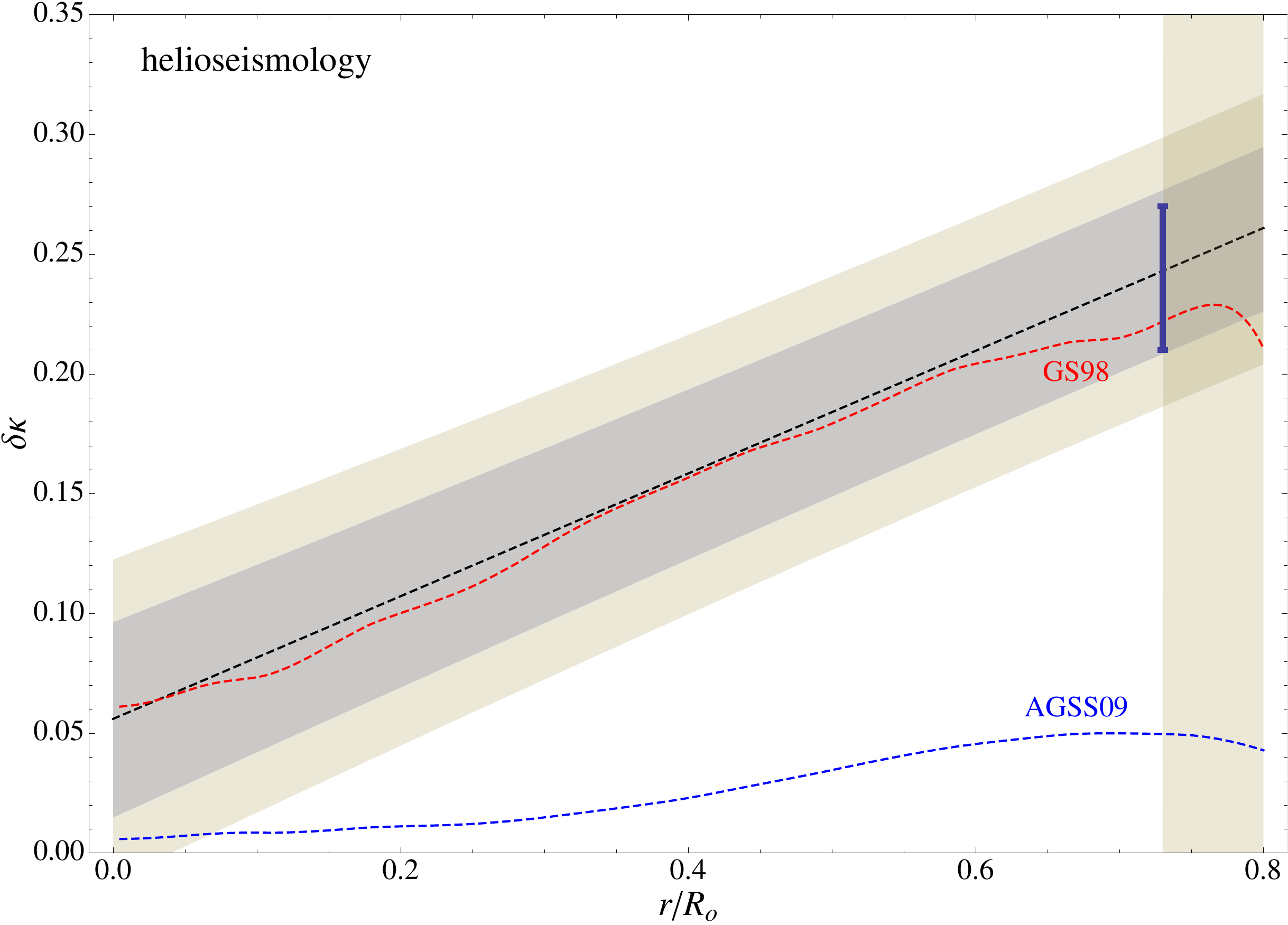}
\includegraphics[width=8.5cm,angle=0]{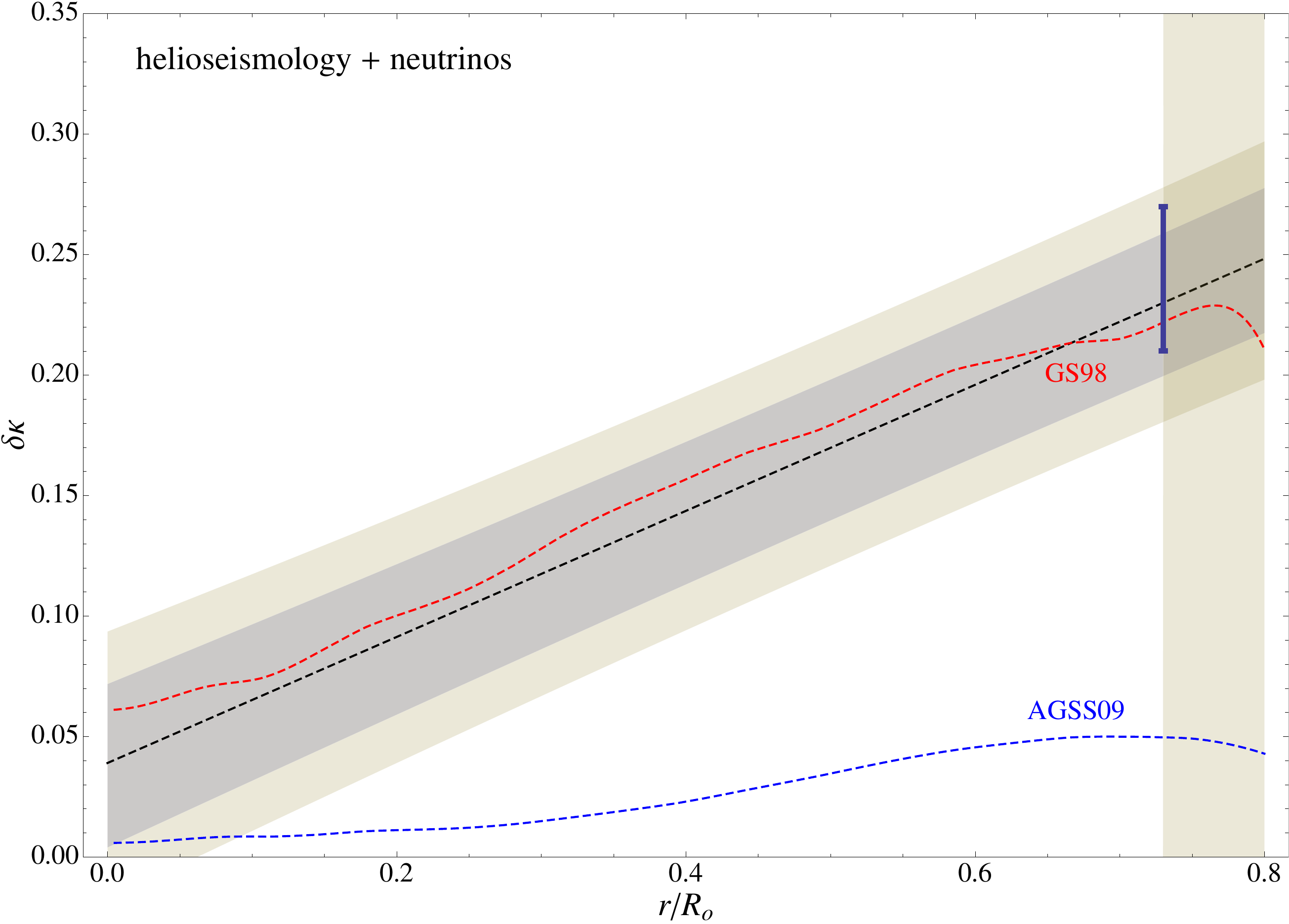}
\end{center}
\par
\vspace{-5mm} \caption{\em {\protect\small Left Panel: The constraints on the opacity profile $\delta \kappa(r)$ obtained from helioseismic data.
 Right panel: The constraints on the opacity profile $\delta \kappa(r)$ obtained from helioseismic and solar neutrino data. See text for details.}}
\label{FigFinal}
\end{figure}

We, finally,  discuss the constraints provided by solar neutrinos on opacity changes parametrized by Eqs.~(\ref{twozones},\ref{lineartilt}).
We note that the coefficients $A_{\rm in} $ and $A_{\rm out} $ and/or 
$A_{0}$ and $A_{1}$ required to fit helioseismic results  (see the previous section), produce enhancements
of the $^8{\rm B}$ and $^7{\rm Be}$ 
neutrino fluxes  which are equal to $28\%$ and $14\%$, respectively. 
While the $^7$Be component would be consistent with the observational 
data, the $^8$B neutrino flux is too large with respect to
the SNO measurement.
When we fit simultaneously helioseismic and solar neutrino data, we obtain 
a slight reduction of the required opacity change. A simple $\chi^2$-analysis gives
$A_{\rm in} = 0.05\pm 0.03$, $A_{\rm out} =0.19\pm 0.03$ for the parameterization (\ref{twozones})
and $A_{0}=0.038\pm 0.034$, $A_{1}=0.192\pm 0.023$ for the parameterization (\ref{lineartilt})
with the best fit values corresponding to $\chi^2_{\rm min}/{\rm d.o.f.} = 2.1 /2$ 
and $\chi^2_{\rm min}/{\rm d.o.f.} = 1.7/2$, respectively. 
The corresponding bounds on $\delta \kappa(r)$ are displayed in fig.~\ref{FigFinal}.


\section{Summary}
\label{sectVIII}

 In this paper, motivated by the solar composition problem and by using the recently developed 
LSM approach \cite{noi}, we have discussed the effects of arbitrary opacity changes on the sun.
Our main results can be summarized as it follow: 
\par\noindent
{\em i) } We have discussed the relation between the effects produced by a variation of the heavy element admixture 
and those produced by a modification of the radiative opacity. We have shown that the relevant quantity
is the variation of the opacity profile $\delta \kappa(r)$ defined in eq.(\ref{kappasource}), that is approximately  
given by the superposition of the {\em intrinsic} opacity change $\delta \kappa_{\rm I}(r)$ and 
the {\em composition} opacity change $\delta \kappa_{\rm Z}(r)$, 
defined in eqs.(\ref{kappaintrinsic}) and (\ref{kappacomposition}) respectively.
\par\noindent 
{\em ii)} We have studied the response of the sun to an arbitrary modification of the opacity $\delta  \kappa(r)$. Namely,
we have calculated numerically the kernels that, in a linear approximation, relate the opacity change 
$\delta \kappa(r)$ to the corresponding modifications of the solar observable properties. 
We have considered the following observable quantities: the squared isothermal sound speed $u(r)$;
the surface helium abundance $Y_{\rm b}$; the convective radius $R_{\rm b}$; the solar neutrino fluxes $\Phi_{\nu}$.
\par\noindent
{\em iii)} We have shown that different observable quantities probe different regions of the sun. 
Moreover, effects produced by variations of opacity in distinct zones of the sun may compensate among 
each other. In this respect, we noted that {\em the sound speed $u(r)$ and the depth of the convective 
envelope $R_{\rm b}$ are practically insensitive to a global rescaling of the opacity.}
\par\noindent 
{\em iv)} As a consequence of the above result, we have seen that the discrepancy between the 
SSM predictions for $u(r)$ and $R_{\rm b}$ and the helioseismic inferred values
can be equally solved by a $\sim 15\%$ decrease of the opacity a the center of the sun 
or by a $\sim 15 \%$ increase of the opacity in the external radiative region.
The degeneracy between these two possible solutions is broken by the ``orthogonal''
information provided by the measurements of the surface helium abundance and of the
boron and beryllium neutrino fluxes, that fix the scale of opacity and indicate that
only the second possibility can effectively solve the solar composition problem.
\par\noindent
{\em v)} We have derived a model-independent relation, eq.~(\ref{dkappab}), that allows to 
obtain a direct determination of the opacity at the bottom of the convective envelope 
$\delta \kappa_{\rm b}$ from quantities that are determined by helioseimic observations, i.e.
the surface helium abundance $\Delta Y_{\rm b}$, the convective radius 
$\delta R_{\rm b}$ and the density at the bottom of the convective region $\delta \rho_{\rm b}$.
By considering the present observation values, we have obtained $\delta \kappa_{\rm b} = 0.24 \pm 0.03$.

 A conclusive view of the present constraints on $\delta \kappa(r)$ is contained in Fig.~\ref{FigFinal}.
The red (blue) dashed lines correspond to the composition opacity changes obtained when we 
replace the AGS05 composition with the GS98 (AGSS09) admixture. 
The blue tick at $r=\overline{R}_{\rm b}$ is the value of the opacity at the bottom of the 
convective region that is obtained by applying the model-independent rel.(\ref{dkappab}) 
to helioseismic data.
The dark and light areas individuate the opacity changes $\delta \kappa(r)$ 
that, for each value of $r$, are obtained at 68.3\% ($1\sigma$) and 95.4\% ($2\sigma$) C.L. by applying a simple 
$\chi^2$ analysis to the opacity modification parametrized by eq.(\ref{lineartilt}).
The left panel takes into account only helioseismic observables. Namely, it is 
obtained by using the present observational determinations of the surface helium abundance and of the 
convective radius. The selected opacity profile $\delta \kappa(r)$ is also consistent with the helioseismic determination 
of the sound speed, as it is has been discussed in fig.~\ref{SoundSpeed2}.
The right panel includes also the observational information on boron and beryllium neutrinos. These data move the required 
opacity change $\delta \kappa(r)$ towards slightly smaller values.
In summary, we see that the helioseismic and solar neutrino data select an opacity profile that 
well corresponds to the GS98 heavy element admixture.
They clearly disfavour different solutions of the solar composition problem, like e.g. that recently proposed by \cite{WimpsSarkar}. 

\section*{Acknowledgments}

It is a pleasure for me to thank B.~Ricci for collaboration on the subject presented in this paper,
for precious suggestions and very useful comments.

\newpage

\section*{Appendix A: The total variation of opacity $\delta \kappa_{\rm tot}(r)$}

In order to calculate the total variation of opacity $\delta \kappa_{\rm tot}(r)$ at a given radius $r$, 
we have to take into account that  the perturbed sun has  different temperature, 
density and chemical composition profiles with respect to SSM. We define:
\begin{equation}
\delta \kappa^{\rm tot}(r) = \frac{\kappa(\rho(r),T(r),Y(r),Z_{\rm i} (r))}
{\overline{\kappa}(\overline{\rho}(r),\overline{T}(r),\overline{Y}(r),\overline{Z}_{\rm i} (r))  } -1
\end{equation}
and, by expanding the above equation to linear order, we obtain:
\begin{equation}
\delta \kappa^{\rm tot}(r) = \kappa_{\rm T} \, \delta T(r) + \kappa_{\rm \rho} \, \delta \rho(r) + \kappa_{\rm Y} \, \Delta Y(r) +
\sum_{\rm i} \kappa_{\rm i} \, \delta Z_{\rm i} (r) +  \delta \kappa_{\rm I}(r)
\label{dkappatot}
\end{equation} 
where:
\begin{eqnarray}
\nonumber
 \kappa_{\rm T}(r) &=& \left. \frac{\partial \ln \kappa}{\partial \ln T} \right|_{\rm SSM}\;\;\;\;\;\;\;\;\;\;\;\;\;\;\;\;\;\; 
\nonumber 
\kappa_{\rho} (r) = \left. \frac{\partial \ln \kappa}{\partial \ln \rho}\right|_{\rm SSM}\\ 
\nonumber
\kappa_{Y}(r) &=& \left.  \frac{\partial \ln \kappa}{\partial Y}\right|_{\rm SSM} \;\;\;\;\;\;\;\;\;\;\;\;\;\;\;\;\;\;\;  
\kappa_{\rm i}(r) =  \left. \frac{\partial \ln \kappa}{\partial \ln Z_{\rm i}}\right|_{\rm SSM}
\end{eqnarray}
and the symbol $|_{\rm SSM}$ indicates that we that we calculate the derivatives $\kappa_{j}(r)$ along the density, temperature and chemical
composition profiles predicted by the SSM.
The quantity $\delta \kappa_{\rm I}(r)$ represents the {\em intrinsic opacity change} and it is given by:
\begin{equation}
\delta \kappa_{\rm I}(r) = \frac{\kappa (\overline{\rho}(r),\overline{T}(r),\overline{Y}(r),\overline{Z}_{\rm i} (r)) }
{\overline{\kappa} (\overline{\rho}(r),\overline{T}(r),\overline{Y}(r),\overline{Z}_{\rm i} (r))  } -1
\end{equation}
By taking advantage of the equation of state of the stellar plasma, we can eliminate the density from Eq.~(\ref{dkappatot}).
We use the relation:
\begin{equation}
\delta \rho(r) = \delta P(r) - \delta T(r) - P_{Y}(r) \Delta Y(r)
\end{equation}
where $P_{Y}(r) \simeq - \partial \ln \mu/\partial Y \simeq - 5/ [8 -5 Y(r)]$ and $\mu$ represents
 the mean molecular weight, and we obtain:
\begin{equation}
\delta \kappa^{\rm tot}(r) = 
\left(\kappa_{\rm T} -\kappa_{\rho}\right)\, \delta T(r) + 
\kappa_{\rm \rho} \, \delta P(r) +
\left(\kappa_{Y} - P_{Y} \, \kappa_{\rho}\right) \, \Delta Y(r) +
\sum_{\rm i} \kappa_{\rm i} \, \delta Z_{\rm i} (r) +  \delta \kappa_{\rm I}(r)
\label{kappatot}
\end{equation} 
In order to evaluate eq.(\ref{kappatot}), we need to estimate the chemical composition of the perturbed sun, i.e.
the quantities $\delta Z_{\rm i}(r)$ and $\Delta Y(r)$. 
By using the approximate method introduced in ref.\cite{noi} and reviewed in the next 
section, we can relate the chemical composition profiles to the 
modification of the photospheric heavy element admixture,
to the present values of $\delta T(r)$
and $\delta P(r)$ and to the parameters $\Delta Y_{\rm ini}$ and $\delta C$ which represent the absolute 
variation of the initial helium abundance and the fractional variation of the pressure at the bottom 
of the convective region, respectively. We obtain:
\begin{equation}
\delta \kappa^{\rm tot}(r) = 
\kappa'_{\rm T} \, \delta T(r) + 
\kappa'_{\rm P} \, \delta P(r) +
\kappa'_{Y} \, \Delta Y_{\rm ini} +
\kappa_{\rm C} \, \delta C +  
\left[ \delta \kappa_{\rm I}(r) 
+\delta \kappa_{\rm Z}(r) \right]
\label{dkappatot2}
\end{equation} 
where $\delta \kappa_{\rm Z}(r)$ is the {\em composition opacity change} given by:
\begin{equation}
\delta \kappa_{\rm Z} (r) = \sum_{i} \kappa_{\rm i} \, \delta  z_{\rm i}
\end{equation}
where $\delta z_{\rm i}$ is the fractional variation of $z_{\rm i}= Z_{\rm i,b}/ X_{\rm b}$, while:
\begin{eqnarray}
\nonumber
\kappa'_{\rm T} &=& \kappa_{\rm T} -\kappa_{\rho}\left( 1+P_{Y} \, \xi_{\rm T}\right) -K_{Y} \,\xi_{\rm T} \\
\nonumber
\kappa'_{\rm P} &=& \kappa_{\rho}\left( 1 - P_{Y} \, \xi_{\rm P}\right) -K_{Y} \,\xi_{\rm P} \\
\nonumber
\kappa_{\rm Y} &=& \left(\kappa_{Y} - \kappa_\rho\, P_{Y}\right) \, \xi_{Y} + Q_{\rm Y} \, \kappa_{\rm Z}\\
\nonumber
\kappa_{\rm C} &=& Q_{\rm C} \, k_{\rm Z}
\end{eqnarray}
with the coefficients $Q_{\rm h}$ and $\xi_{\rm h}$ defined in the next section.
In the derivation of the above equation, we took into account that $\sum_{i} \kappa_{i} = \kappa_{\rm Z}$ where 
$\kappa_{\rm Z} = \partial \ln \overline{\kappa}/\partial \ln Z$ is the partial derivative of opacity with respect to the total 
metal abundance (i.e. calculated by rescaling all the heavy 
element abundances by a constant factor, so that the metal admixture remains fixed).

It is useful to define the opacity change $\delta \kappa(r)$ given by:
\begin{equation}
\delta \kappa(r) = \delta \kappa_{\rm I}(r) + \delta \kappa_{\rm Z}(r)
\end{equation}
which groups together the contributions to $\delta \kappa^{\rm tot} (r)$ which are directly
related to the variation of the input parameters. 
While the other terms in 
eq.(\ref{dkappatot2})  represent {\em derived } quantities that are determined by solving the structure equations and/or by 
fitting the observed properties of the sun, the opacity change $\delta \kappa(r)$
can be varied, in principle, in arbitrary way.

\section*{Appendix B: The chemical composition of the sun}

The chemical composition of the perturbed sun should be calculated by integrating the 
the perturbed structure and chemical-evolution equations starting from an ad-hoc chemical homogeneous ZAMS model. 
In ref.\cite{noi}, we proposed a simplified approximate procedure that allows to estimate with sufficient
 accuracy the helium and metal abundances of the modified sun, without requiring to 
follow explicitly its time-evolution. We review this procedure and we extend it 
to take into account the effect of a variation of the photospheric composition.

In order to quantify the relevance of the different mechanisms determining the
present composition of the sun, we express the helium and metal abundance
according to:
\begin{eqnarray}
\nonumber
Y(r)&=&Y_{\rm ini}\,\left[ 1+D_{Y}(r) \right] +Y_{\rm nuc}(r)\\
Z_i (r)&=&Z_{\rm i, ini}\,\left[ 1+D_{Z}(r) \right]
\end{eqnarray}
Here,  $Y_{\rm ini}$ and $Z_{\rm i, ini}$ are the initial values for the abundances,
the terms $D_{Y}(r)$ and $D_{Z}(r)$ describe the effects of elemental diffusion
and $Y_{\rm nuc}(r)$ represents the total amount of helium produced in
the shell $r$ by nuclear processes. We note that we, implicitly, assumed that heavy elements
have all the same diffusion velocity by introducing a common diffusion term 
$D_{Z}(r)$ for all metals.

We are interested in describing how the chemical composition is modified when we perturb
the SSM.  In the radiative core ($r\le \overline{R}_{\rm b}$), we neglect the effects produced 
by variations of the diffusion terms\footnote{The diffusion terms $D_{Y}(r)$ and $D_{Z}(r)$ are at the few per cent level
in the radiative region. Their variations are, thus, expected to produce very small effects on the solar composition.} 
and we write:
\begin{eqnarray}
\nonumber
\Delta Y(r) &=& \Delta Y_{\rm ini} \,\left[ 1+\overline{D}_{Y}(r) \right] +\Delta Y_{\rm nuc}(r) \\
\delta Z_{\rm i}(r) &=& \delta Z_{\rm i, ini}
\label{ChimRad}
\end{eqnarray}
where $\Delta Y_{\rm nuc}(r)$ is the absolute variation of the amount of helium produced by
nuclear reactions. A better accuracy is required in the convective region, because the surface helium abundance $Y_{\rm b}$ 
is an observable quantity. We, thus, discuss explicitly the role of diffusion and we write:
\begin{eqnarray}
\nonumber
\Delta Y_{\rm b}&=& (1+\overline{D}_{Y, \rm b})\, \Delta Y_{\rm ini}  + \overline{Y}_{\rm ini}\, \overline{D}_{Y,\rm b} \, \delta D_{Y,{\rm b}}\\
\delta Z_{\rm i, b}&=&\delta Z_{\rm i, ini} +\frac{\overline{D}_{Z, \rm b}}{1+\overline{D}_{Z, \rm b}}\, \delta D_{Z,_{\rm b}}
\label{ChimSup}
\end{eqnarray}
where $\delta D_{Y,\rm b}$  and $\delta D_{Z, \rm b}$ are the fractional variations of the diffusion terms
$D_{Y,\rm b}$ and $D_{Z, \rm b}$.
The quantities $\Delta Y_{\rm b}$ and $\delta Z_{\rm b}$
are related among each other, since the metals-to-hydrogen ratios at the
surface of the sun are observationally fixed.  If we indicate with $z_{\rm i} = Z_{\rm i, b}/X_{\rm b}$ 
the surface abundance of the $i-$element (rescaled to that of hydrogen), we obtain:
\begin{equation}
\delta Z_{\rm i, b}=-\frac{1}{1-\overline{Y}_{\rm b}}\Delta Y_{\rm b} + \delta z_{\rm i}
\label{Zsup}
\end{equation}
where we considered that $X_{\rm b} \simeq 1-Y_{\rm b}$, while $\delta z_{\rm i}$ is defined by:
\begin{equation}
\delta z_{\rm i} = \frac{(Z_{\rm i,b}/X_{\rm b}) -(\overline{Z}_{\rm i, b}/\overline{X}_{\rm b})} {(\overline{Z}_{\rm i, b}/\overline{X}_{\rm b})}
\end{equation}
 The above relation can be rewritten in terms of the initial helium and metal abundances,
obtaining:
 \begin{equation}
\delta Z_{\rm i, ini} = Q_0 \, \Delta Y_{\rm ini}
+Q_1 \, \delta D_{Y,{\rm b}}
 +Q_2 \, \delta D_{Z,_{\rm b}} + \delta z_{\rm i}
\label{Zini}
\end{equation}
The coefficients $Q_{i}$ have been calculated explicitly in ref.\cite{noi} 
and are given by $Q_0  = - 1.141 $, $Q_1  = 0.041$ and $Q_2 = +0.118$, respectively.

 Up to this point, the derived relations have a general validity, since the only assumption implied by our analysis 
is that the heavy elements have all the same diffusion velocity (we take iron as representative for all metals). 
To complete our calculation, we have to estimate the term $\Delta Y_{\rm nuc}(r)$ in eq.(\ref{ChimRad}) and the quantities 
$\delta D_{Y,{\rm b}}$ and  $\delta D_{Z,{\rm b}}$ in eqs.(\ref{ChimSup}, \ref{Zini}). We use the procedure adopted in the LSM approach, 
where we assumed that the helium produced by nuclear reactions scales proportionally to the energy generation 
coefficient (and, thus, the helium production rate) in the present sun, i.e: 
\begin{equation}
\Delta Y_{\rm nuc} = \overline{Y}_{\rm nuc} (r) \, \delta \epsilon^{\rm tot}(r) 
\end{equation}
where $\delta \epsilon^{\rm tot}(r)$ is the fractional variation of the energy generation rate. 
The effect of elemental diffusion is modelled by assuming that the terms $D_{\rm i,b}$ vary 
proportionally to the efficiency of diffusion in the present sun, obtaining (see sect.6.3 and appendix C of ref.\cite{noi}):
\begin{eqnarray}
\nonumber
\delta D_{Y,\rm b} &=& \Pi_Y \, \delta T_{\rm b} + \Pi_{P} \, \delta P_{\rm b}\\
\delta D_{Z,\rm b} &=& \Pi_Z \, \delta T_{\rm b} + \Pi_{P} \, \delta P_{\rm b}
\label{diff}
\end{eqnarray}
where $\Pi_Y =2.05$, $\Pi_Z =2.73$ and $\Pi_P = -1.10$.

By following the calculations described in sect. 6.2 of \cite{noi},  we obtain the following expression 
for the variation of the helium abundance in the radiative region:
\begin{equation}
\Delta Y(r) = \xi_{Y}(r) \, \Delta Y_{\rm ini} + \xi_T(r) \, \delta T(r) + \xi_{P}(r) \, \delta P(r) 
\end{equation}
The coefficients  $\xi_{h}(R)$ are defined in eq.(31) of ref.\cite{noi} and are shown in their fig.~4.

By taking into account relations (\ref{diff}) and by considering the conditions that hold at the 
bottom of the convective region (expressed in eq.~(21) of \cite{noi}), we can estimate the variation of 
metal abundances in the radiative region, obtaining:
\begin{equation}
\delta Z_{\rm i} (r) = \delta Z_{\rm i, ini} = Q_{Y} \, \Delta Y_{\rm ini} + Q_{C} \, \delta C + \delta z_{\rm i}  
\end{equation}
where $Q_{\rm Y} =-0.887$, $Q_{C} =-0.164$ and $\delta C= \delta P_{\rm b}$ represents the variation of pressure at the 
bottom of the convective envelope.

Finally, we can calculate the abundances in the convective region obtaining
\begin{eqnarray}
\delta Y_{\rm b} &=&  A_{Y} \, \Delta Y_{\rm ini} + A_{C} \, \delta C  \\
\delta Z_{\rm i, b} &=&  B_{Y} \, \Delta Y_{\rm ini} + B_{C} \, \delta C + \delta z_{\rm i} 
\end{eqnarray}
where $A_{Y}=0.838$, $A_{C}=0.033$, $B_{Y}=-1.088$ and $B_{C}=-0.043$.

We remark that, while the quantities $\Delta Y_{\rm ini}$ and $\delta C$ are  
parameters which are univocally determined by imposing the appropriate integration conditions (see next section),
the quantities $\delta z_{\rm i}$ represent {\em input parameters} for solar model calculations.

\section*{Appendix C: Linear Solar Models}

By expanding to linear order the structure equations of the present sun close 
to the SSM solution and by assuming that the variation of the chemical abundances of the 
sun can be estimated by the procedure outlined in the previous section,
we obtain a linear system of ordinary differential equations that completely determine the physical 
and chemical properties of the ``perturbed'' sun (see \cite{noi} for details).
Namely, we obtain:
\begin{eqnarray}
\frac{d\delta m}{dr} &=& \frac{1}{l_m} \,\left[ \gamma_{P}\, \delta P + \gamma_T \, \delta T - \delta m 
+ \gamma_Y \, \Delta Y_{\rm ini}  \right]\label{linsyst3}\\
\nonumber
\frac{d\delta P}{dr}&=& \frac{1}{l_P}\, \left[\left(\gamma_{P}-1\right) \, \delta P + \gamma_T \, \delta T + \delta m  
+ \gamma_Y \, \Delta Y_{\rm ini}  \right]\\
\nonumber
\frac{d\delta l}{dr}& = & \frac{1}{l_l}\, \left[ \beta'_P \,\delta P +  \beta'_T \,\delta T - \delta l + 
 \beta'_Y \,\Delta Y_{\rm ini} + \beta'_C \, \delta C  \right]\\
\nonumber
\frac{d\delta T}{dr}&=& \frac{1}{l_T}\, \left[ \alpha'_P \,\delta P + \alpha'_T \,\delta T + \delta l + 
\alpha'_Y \, \Delta Y_{\rm ini}+ \alpha'_C \, \delta C + \delta \kappa  \right] 
\label{lsm}
\end{eqnarray}
The coefficients $\gamma_{h}$, $\beta'_{h}$ and $\alpha'_{h}$ and the scale heights $l_{h}\equiv \left[d\ln( \overline{h})/dr\right]^{-1} $ have been
calculated in \cite{noi} and are shown in their fig.~1, fig.~5 and fig.~6. The parameters $\Delta Y_{\rm ini}$ and $\delta C$ represent 
the absolute variation of the initial helium abundance and the relative variation of pressure at the
bottom of the convective envelope and can be univocally determined by imposing the appropriate integration conditions.  
At the center of the sun ($r=0$) we have:
\begin{eqnarray}
\nonumber
\delta m &=& \gamma_{P,0}\, \delta P_{0} + \gamma_{T,0} \, \delta T_{0} 
+ \gamma_{Y,0} \, \Delta Y_{\rm ini}  \;\;\;\;\;\;\;\; \;\;\;\;\;\;\;\;\;\;\;\;\;\;\;\;\;\;\;\;\;\;\;\;\;\;\;\;\;
\delta P = \delta P_{0}\\
\nonumber
\delta l &=& \beta'_{P,0} \,\delta P_{0} +  \beta'_{T,0} \,\delta T_{0} + 
 \beta'_{Y,0} \,\Delta Y_{\rm ini} + \beta'_{C,0} \, \delta C   \;\;\;\;\;\;\;\; \;\;\;\;\;\;\;\;\;\;\;\;\;
\delta T = \delta T_{0} 
\end{eqnarray}
where the subscript "0" indicates that a given quantity is evaluated at $r=0$. At the bottom of the convective envelope ($r=\overline{R}_{\rm b}$), 
we have instead:
\begin{eqnarray}
\nonumber
\delta m &=& - \overline{m}_{\rm conv} \, \delta C \;\;\;\;\;\;\;\; \;\;\;\;\;\;\;\;\;\;\;\;\;\;\;\;\;\;\;\;\;\;\;
\delta P  =  \delta C\\
\nonumber
\delta l &=& 0      \;\;\;\;\;\;\;\;\;\;\;\;\;\;\;\;\;\;\;\;\;\;\;\;\;\;\;\;\;\;\;\;\;\;\;\;\;\;\;\;\;\;\;\;\;\;
\delta T = A'_Y \, \Delta Y_{\rm ini} + A'_{C} \, \delta C
\end{eqnarray}
where $A'_Y=0.626$ and $A'_C=0.025$ and $\overline{m}_{\rm conv}=\overline{M}_{\rm conv}/M_{\odot}=0.0192$ is the fraction of solar mass contained in the convective region.

The term $\delta \kappa(r)$ contains the contributions to the modification of the opacity profile of the sun that are directly related to 
the variation of the input parameters. It is given by:
\begin{equation}
\delta \kappa(r) = \delta \kappa_{\rm I}(r) + \delta \kappa_{\rm z}(r)
\end{equation}
where $\delta \kappa_{\rm I}(r)$ and $\delta \kappa_{\rm Z}(r)$ are the {\em intrinsic} and {\em composition} opacity changes, 
defined in eqs.(\ref{kappaintrinsic}) and (\ref{kappacomposition}) respectively.
It represents the source term that drives the modification of the solar properties an that can be bounded by observational data.

\newpage

\section*{\sf  References}
\def\refname{

\end{document}